\documentclass[twocolumn, twocolappendix]{aastex631}

\usepackage{float}
\usepackage{bm}
\usepackage[multiple]{footmisc}
\begin{document}

\newcommand{\newtext}[1]{}

\title{Counting the Unseen I: Nuclear Density Scaling Relations for Nucleated Galaxies}

\author[0000-0002-7064-3867]{Christian H. Hannah}
\affiliation{Department of Physics and Astronomy, University of Utah\\ 115 South 1400 East, Salt Lake City, Utah 84112, USA}

\author[0000-0003-0248-5470]{Anil C. Seth}
\affiliation{Department of Physics and Astronomy, University of Utah\\ 115 South 1400 East, Salt Lake City, Utah 84112, USA}

\author[0000-0002-4337-9458]{Nicholas C. Stone}
\affiliation{Racah Institute of Physics, The Hebrew University, 91904, Jerusalem, Israel}

\author[0000-0002-3859-8074]{Sjoert van Velzen}
\affiliation{Leiden Observatory, Leiden University, Postbus 9513, 2300 RA, Leiden, The Netherlands}

\begin{abstract}

The volumetric rate of tidal disruption events (TDEs) encodes information on the still-unknown demographics of central massive black holes (MBHs) in low-mass galaxies ($\lesssim 10^9$~M$_\odot$). Theoretical TDE rates from model galaxy samples can extract this information, but this requires accurately defining the nuclear stellar density structures. This region is typically dominated by nuclear star clusters (NSCs), which have been shown to increase TDE rates by orders of magnitude. Thus, we assemble the largest available sample of pc-scale 3-D density profiles \newtext{that include NSC components}. We deproject the PSF-deconvolved surface brightness profiles of 91 nearby galaxies of varying morphology and combine these with nuclear mass-to-light ratios estimated from measured colors or spectral synthesis to create 3-D mass density profiles. We fit the inner 3-D density profile to find the best-fit power-law density profile in each galaxy. We compile this information as a function of galaxy stellar mass to fit new empirical density scaling relations. These fits reveal positive correlations between galaxy stellar mass and central stellar density in both early- and late-type galaxies.  We find that early-type galaxies have somewhat higher densities and shallower profiles relative to late-type galaxies at the same mass. We also use the density profiles to estimate the influence radius of each galaxy's MBH and find that the sphere of influence was likely resolved in most cases. These new relations will be used in future works to build mock galaxy samples for dynamical TDE rate calculations, with the aim of constraining MBH demographics in low-mass galaxies.

\end{abstract}


\section{Introduction} \label{sec:introduction}
Massive black holes (MBHs) have been found at the centers of most galaxies, and these objects are believed to significantly influence the evolution of their host galaxies \citep[e.g.][]{Kormendy2013}. Although their existence has become increasingly clear, the initial formation of MBHs and the details of their subsequent growth are still unclear. 

Several theories have been proposed for the seed formation mechanisms of MBHs \citep[see][for recent reviews]{Inayoshi2020,Greene2020}. First, MBHs may form from the remnants of the first stars. Stars born from the theoretically expected pristine (metal-free) gas in the early universe are predicted to be extremely massive \citep[Pop-III stars, e.g.][]{Bromm2004,Yoshida2006,Hirano2014} and could produce $\gtrsim$100~M$_\odot$ black holes at the end of their lives \citep[e.g.][]{Fryer2001,Bromm2003,Spera2017}. A second scenario, known as direct collapse, involves extremely massive clouds of gas collapsing directly into a black hole seed, bypassing the standard phases of stellar evolution \citep[e.g.][]{Loeb1994,Johnson2011,Agarwal2012, Dunn2018,Wise2019}. This mechanism is also enabled by the pristine gas at early times and is believed to result in much larger seed black holes \citep[$\sim10^4 - 10^6$~M$_\odot$, e.g.][]{Begelman2006,Lodato2006,Begelman2008}. Lastly, the gravitational runaway scenario explains seed black hole formation via repeated collisions of massive stars and black holes within dense star clusters \citep[e.g.][]{Portegies-Zwart2002,Devecchi2009, Stone2017,Shi2021,Rizzuto2021}. The final outcome of this seeding mechanism varies substantially with initial conditions, resulting in a wide distribution of potential seed MBH masses. 

Each of these theories is compatible with observed MBH demographics in high-mass galaxies, where subsequent growth through active galactic nucleus episodes washes out memory of initial seed mass distributions \citep{Soltan1982}; however, they produce observable differences in the masses and occupation fraction of MBHs in low-mass ($\lesssim 10^9$~M$_\odot$) galaxies \citep[e.g.][]{Volonteri2008,Ricarte2018,Greene2020}. Dynamical detections of the lower-mass MBHs expected in these galaxies are extremely difficult \citep[e.g.][]{Nguyen2018,Zocchi2019}. It is also challenging to measure black hole accretion in low-mass galaxies due to the luminosity dependence on the MBH mass and the presence of contaminants (e.g. X-ray binaries) at lower luminosities \citep[e.g.][]{Miller2012,Reines2022}. A promising approach to constraining MBH demographics in this regime is through observations of tidal disruption events (TDEs) and the rates at which they occur as a function of galaxy mass.

TDEs occur when a star passes close enough to a massive black hole that tidal forces overcome the star’s self-gravity, resulting in the destruction of the star and the formation of an accretion flow \citep{Rees1988}.
Using these transient events to inform MBH demographics has been limited in the past by low number statistics \citep[$\sim$100 detected to date, e.g.][]{Gezari2021}. The Rubin Observatory, which will come online in 2025, should greatly improve the statistics issue, as it is expected to detect more than 10 of these events per night \citep{Bricman2020}. A similar detection rate is expected for the wide-field UV survey satellite {\it ULTRASAT} \citep{Ben-Ami2023}.  However, translating these anticipated TDE detections into constraints on the MBH occupation fraction or mass function requires accurate TDE rate predictions over the range of galaxy properties where the events are detected. This need for accurate TDE rate predictions is the primary motivation for this work.

TDE rates are set by the evolution of stellar orbits into radial, low-angular momentum configurations defined by a ``loss cone'' in velocity space \citep{Frank1976, Cohn1978}.  Near the central MBH, the loss cone is devoid of stars, and TDE rates are ultimately determined by how quickly relaxational processes, such as two-body scattering, can diffuse stars through phase space into the loss cone.  Since TDE rates are set by relaxation times, they are sensitive to the stellar distribution around the MBH, and different stellar density profiles can produce radically different rates of disruption. TDE rates are dominated by the densities of stars on parsec scales, but at the distances where TDE flares are detected, stellar densities can at best be measured on scales $\sim 10$ times larger \citep{French2020a}. Consequently, most theoretical estimates of TDE rates calibrate the stellar distributions from resolved observations of very nearby galactic nuclei \citep{Magorrian1999, Wang2004, StoneMetzger2016, Stone2016}.

\citet{Pfister2020} have shown that in low-mass galaxies, the TDE rate is completely dominated by the presence of nuclear star clusters (NSCs) at their centers. This effect is further highlighted in \citet{Polkas2023}, who find that TDE rates originating from NSCs are on average 3 orders of magnitude higher than TDE rates produced by the underlying galaxy components (i.e. disk and bulge) across all galaxy masses via cosmological TDE rate simulations that reproduce observed volumetric rates. \newtext{In this paper, we aim to quantify the density profiles of galaxies including their NSC components, which were previously excluded from some TDE rate estimates \citep[e.g.][]{StoneMetzger2016}.}

\newtext{NSCs are dense stellar systems with observed central densities as high as $\sim10^7$~M$_\odot$/pc$^3$ that occupy the central $\sim$10~pc of a galaxy \citep[see][for a recent review]{Neumayer2020}.} \newtext{These objects are detected as excess brightness above an inward extrapolation of the underlying host galaxy light profile.} \newtext{Nearby NSCs have light profiles that are well represented by S\`ersic functions \citep[e.g.][]{Graham2009,Carson2015,Hoyer2023a}.} \newtext{Given their compactness (median effective radii $\sim$5pc), they are often spatially unresolved (and often undetectable) at large distances ($\gtrsim$20~Mpc).} 

The increased TDE rates for NSC host galaxies, paired with the prevalence of nuclear star clusters (NSCs) in galaxies between $10^8$ and $10^{10}$ M$_\odot$ \citep[$\sim$70\%;][]{Sanchez-Janssen2019,Neumayer2020,Hoyer2021}, suggest that future TDE detections from low-mass galaxies will come almost entirely from those harboring NSCs at their centers (also known as nucleated galaxies).  Therefore, model galaxy samples used to estimate volumetric TDE rates should reflect the elevated central densities observed in NSCs.  It is also plausible that other types of nuclear transients generated by relaxational or collisional processes, such as some LIGO-band gravitational wave signals \citep{Antonini2016}, future {\it LISA}-band extreme mass ratio inspirals \citep{Hopman2005, Qunbar2023}, and X-ray quasi-periodic eruptions \citep{Miniutti2019, Wevers2022}, will likewise be dominated by NSC dynamics.

With these motivations in mind, we present the largest available set of high-resolution 3-D stellar density profiles of galaxies hosting NSCs. We use these density profiles to develop relations between host galaxy stellar mass \& type and their central densities \& gradients. 
In \newtext{Papers II \& III} (Hannah et al., {\em in prep}), we will improve upon existing volumetric TDE rate estimates by using these new relations to inform the model galaxy samples used in our rate calculations. 
The galaxy sample used in this work is described in Section \ref{sec:sample}. Section \ref{sec:ml} describes the definition of the mass-to-light ratios for each galaxy, which are crucial for converting photometry to mass density (Section \ref{sec:nuc_dens}). The resultant nuclear density scaling relations are presented in Section \ref{sec:relations}. Lastly, how these scaling relations will be utilized in future TDE rate modeling, and their viability for such modeling is discussed in Sections \ref{sec:soi} and \ref{sec:con-future}.

\section{The Sample} \label{sec:sample}

\subsection{1-D Surface Brightness Profiles}
We selected our galaxy sample from five existing studies of nearby galaxy surface brightness (SB) profiles derived from \textit{Hubble Space Telescope} (\textit{HST}) imaging \citep{Lauer1995,Lauer2005,Pechetti2017,Pechetti2020,Hoyer2023a}. 
\citet{Pechetti2020} and \citet{Hoyer2023a} focus solely on galaxies hosting NSCs, while the samples in \citet{Pechetti2017}, \citet{Lauer1995}, and \citet{Lauer2005} consist of both nucleated and non-nucleated galaxies. Our primary goal is to characterize the \newtext{nuclear} mass profiles of \newtext{NSC-hosts} across both early- and late-type galaxies, however, we characterize some non-nucleated galaxies from these literature sources as well (see Section \ref{sec:division} for more discussion). 

\begin{figure}[t] 
    \centering
    \includegraphics[width=0.45\textwidth]{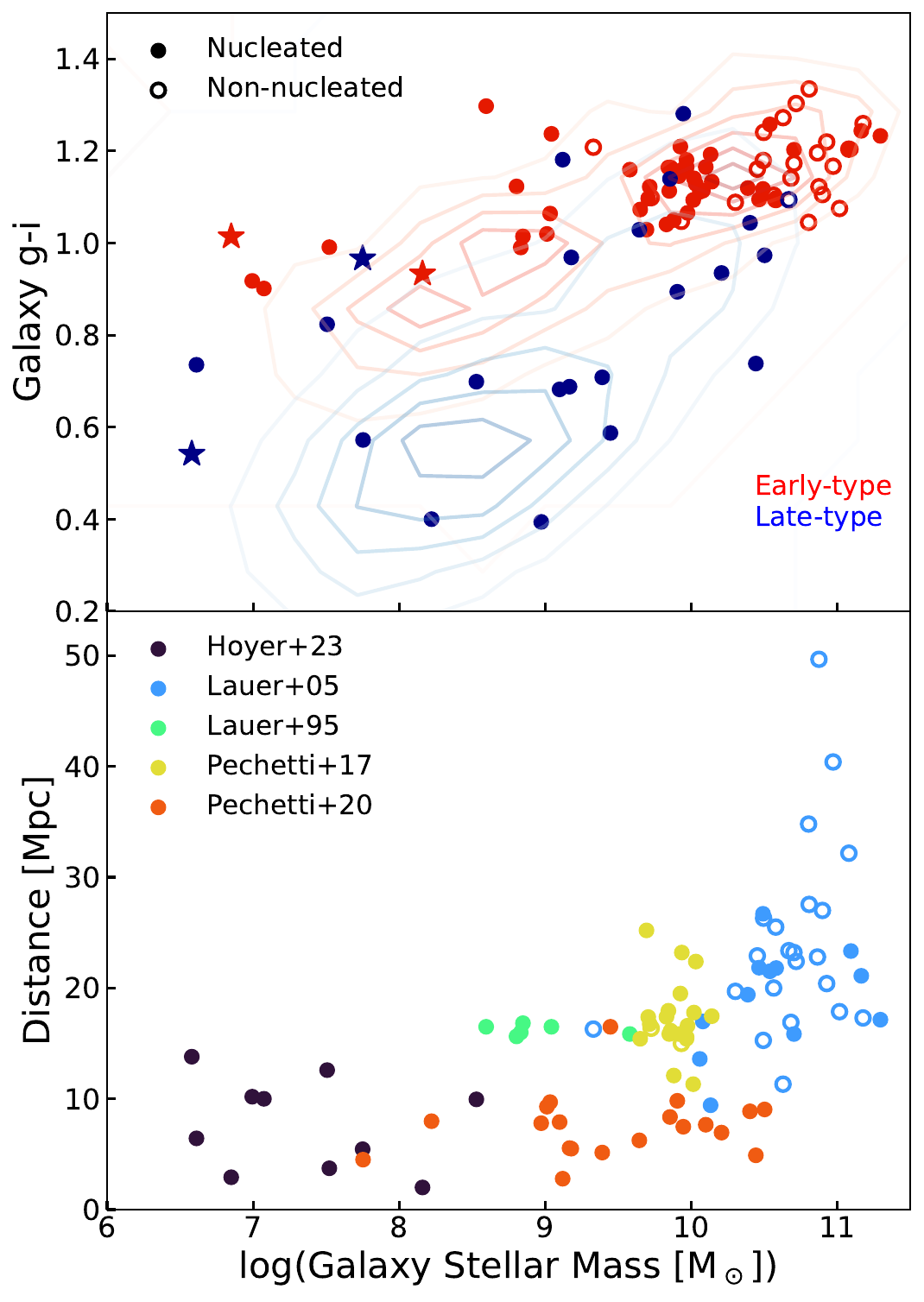} 
    \caption{Top: The color-magnitude diagram of our sample, which highlights the mass coverage and morphological diversity of the sample. The colors in this panel indicate the galaxy type, while closed symbols indicate NSC hosts. Nuclear $g-i$ colors were used for the four galaxies indicated with star symbols due to a lack of existing galaxy color measurements. The shaded contours give the distributions of all galaxies in 50MGC. Bottom: The distance distribution of our galaxies with colors indicating the parent sample. Similarly to the top panel, nucleation is indicated by closed symbols.}
    \label{fig:CMD}
\end{figure}

Overall, our sources contained SB profile measurements for 185 galaxies; however, many galaxies were unsuitable for the analysis here due to several factors. As our aim is to resolve the \newtext{nuclear} density profiles \newtext{on scales resolving the NSC}, the majority of rejections were due to a lack of nuclear spatial resolution. NSCs are generally very compact with median effective radii of $5$~pc \citep{Neumayer2020}. Therefore, to ensure we resolve the NSCs in our galaxies, we require PSF-deconvolved SB measurements with a minimum radius of at least 5~pc. \newtext{At HST resolution, this corresponds to a single pixel at the distance of the Virgo cluster (16~Mpc). The handful of galaxies above this distance limit have SB profile data from \citet{Lauer1995,Lauer2005} to half-pixel resolution recovered from dithering.}

\newtext{\citet{Lauer1995,Lauer2005} use the Lucy-Richardson algorithm \citep{Lucy1974,Richardson1972} to deconvolve their images before measuring the SB profile, while the remaining sources present model fits (S\`ersic profiles) that account for the PSF (which is subsequently deconvolved). Both techniques utilized model PSFs derived from the Tiny Tim package \citep{Krist1995}. We note that for model fits, the PSF can only be accounted for accurately if the data is well represented by the assumed model, which, as mentioned previously, is generally true for NSCs fit with S\`ersic profiles.}

For sources that provide SB data as model fits, we define the minimum radial resolution as the larger of half the image pixel scale or half the point spread function (PSF) full-width half max (FWHM). Given that the \textit{HST} PSF is similar in all optical cameras and filters, we adopt a fiducial PSF FWHM of $0\farcs07$ for most observations. For the two galaxies imaged in IR (WFC3 F110W \& F160W), we use PSF FWHMs of $0\farcs13$ and $0\farcs15$, respectively. Ten of the galaxies with ample spatial resolution were still rejected based on the absence of nuclear color data, which is required to estimate mass-to-light ratios (see Section~\ref{sec:ml}). Three galaxies from \citet{Lauer2005} were removed due to clear PSF deconvolution errors visible in the 1-D SB profiles (IC~1459, NGC~7213, and NGC~4278). Another three galaxies were rejected based on a lack of galaxy stellar mass measurements, making them unsuitable for use in the scaling relations. These cuts result in a sample of 94 (69 nucleated) galaxies where nuclear densities can be accurately measured.

Next, to ensure the nuclear light profiles of these 94 galaxies were dominated by starlight, we examined each for potential AGN contamination. Potential contaminants were first identified using the \citet{Veron-Cetty2010} AGN catalog. Then, to gauge if these AGN are expected to contribute significantly to the optical light profile, we use the $\alpha_{\rm OX}$ power-law relationship for the AGN continuum at optical to X-ray wavelengths \citep{Elvis2002} to estimate the optical luminosities. 
Specifically, the optical AGN luminosities were estimated in the primary photometry band by integrating the power-law SED defined by $\alpha_{\rm OX}$ and normalized by the published X-ray luminosity over the wavelength range covered by the specific filter. Given that the AGN continuum is expected to be point-like, we compare these luminosities to the nuclear galaxy luminosity contained within a radius of $0\farcs081$; this corresponds to the 50\% encircled energy radius for ACS/WFC $F814W$ imaging. We note that this radius does not vary significantly for other filters and instruments, and thus, we use the same aperture for all measurements.

In total, eight of our sample galaxies were found to host AGN (Circinus, NGC~3607, NGC~4472, NGC~4552, NGC~2787, NGC~4941, NGC~4517, NGC~2974). When estimating their optical luminosities, we used published X-ray luminosities from Chandra in \citet{Bi2020} for six of them. For NGC~4517 and NGC~2974, we used X-ray luminosities from \citet{She2017} (Chandra) and \citet{Osullivan2001} (EINSTEIN), respectively. In all but three of the galaxies (NGC~2974, NGC~4941, Circinus), the predicted AGN luminosities were much less than the observed optical fluxes.  We therefore removed those three galaxies but kept the rest.  We note that even for the three bright AGN host galaxies, their nuclear density measurements were typical of other measurements for similar galaxies. Thus, it is unlikely that our other density measurements would be significantly affected by an undetected AGN continuum component. After removing the three bright AGN galaxies, our final sample consists of 91 galaxies (67 nucleated). The median distance of our final galaxy sample is 16.1~Mpc, with the most distant object at 49.7~Mpc.  We provide details on each of the studies we draw from below, including the mass range of the galaxies and their morphologies. 

\textbf{\textit{Hoyer et al. (2023)}} -- 10 galaxies -- SB data is given as PSF convolved S\'ersic fits to the NSCs of \textit{HST} WFC3, ACS and WFPC2 imaging for nearby galaxies. These are all very low-mass galaxies with log($M_{\rm gal}$)~$<8.6$. In this work, the galaxy was not fit directly but was instead modeled as a constant light profile over the nuclear region.
    
\textbf{\textit{Pechetti et al. (2020)}} -- 20 galaxies -- SB data is given as Multi-Gaussian Expansion (MGE) fits to S\'ersic profiles derived from PSF convolved fits to \textit{HST} WFC3, ACS, and WFPC2 data. All galaxies in this sample host NSCs and span a mass range of $7.7<$~log($M_{\rm gal}$)~$<10.5$. 86\% of the galaxies included from this source are late-types.
    
\textbf{\textit{Pechetti et al. (2017)}} -- 21 galaxies -- Similarly to \citet{Pechetti2020}, SB data is given via MGE fits to S\'ersic profiles derived from PSF convolved fits to \textit{HST} WFC3, ACS, and WFPC2 images. The sample includes intermediate-mass ($9.7<$~log($M_{\rm gal}$)~$<10.5$) early-type galaxies. These galaxies were not selected to be nucleated, but 90\% of them do host NSCs (see Section \ref{sec:division} for more details on nucleation designations).  

\textbf{\textit{Lauer et al. (2005)}} -- 34 galaxies -- This source provides PSF deconvolved SB profiles from \textit{HST} WFPC2/PC images for a large sample of early-type galaxies. The sample spans a wide range in galaxy stellar masses and contains both galaxies with and without NSCs. The galaxies we draw from this work have masses within $9.3<$~log($M_{\rm gal}$)~$<11.3$ and 35\% host NSCs. 
    
\textbf{\textit{Lauer et al. (1995)}} -- 6 galaxies -- Similar to \citet{Lauer2005}, but the PSF deconvolved SB profiles come from \textit{HST} WFPC1 imaging of nearby early-types. Due to reliability issues of WFPC1 photometry at radii $<0\farcs1$, we only select the galaxies close enough that $0\farcs1$ corresponds to a physical distance $<10$~pc ($\sim$8~pc for all 6 galaxies). All six galaxies are low-mass (log($M_{\rm gal}$)~$<9.6$) and host NSCs.

\begin{figure}[t]
    \centering
    \includegraphics[width=0.45\textwidth]{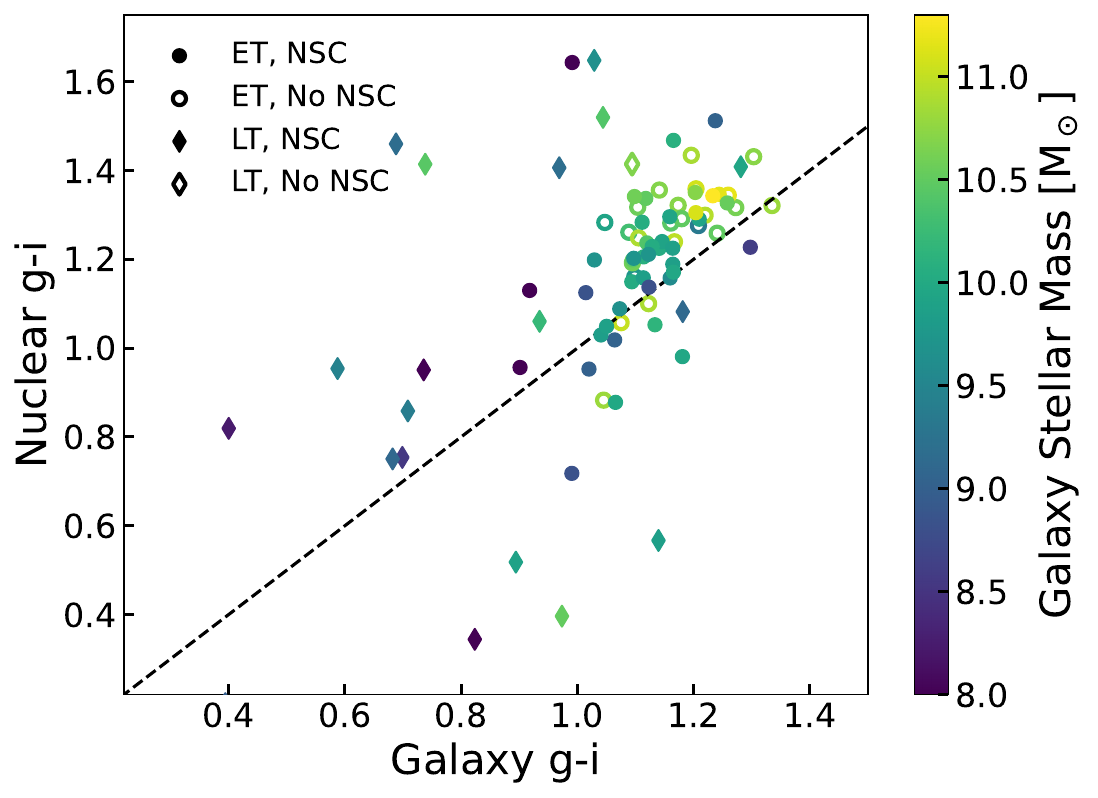}
    \caption{Comparison between nuclear $g-i$ ($r<0\farcs5$) and galaxy-wide $g-i$ with data points colored by the galaxy stellar mass. Early-type galaxies are indicated with bullets, while late-types are shown as diamonds. Nucleation is indicated with closed symbols.}
    \label{fig:gal_v_nuc_gi}
\end{figure}

\subsection{Host Galaxy Properties} \label{sec:host_properties}

Our goal is to expand on the relation between nuclear stellar densities and galaxy stellar mass from \citet{Pechetti2020}, including a separation of galaxies into early- and late-type.  Therefore, we need a consistent set of galaxy stellar mass and morphology measurements. Most of our measurements come from the 50~Mpc galaxy catalog \citep[50MGC;][]{Ohlson2023}. This work provides self-consistent galaxy stellar masses, distances, and morphological designations, as well as galaxy-wide color measurements. The published morphology designations for our sample galaxies (where available) agree with the morphologies listed in the 50MGC. For late-type galaxies with $g-i$ colors along the red sequence in the color-mass diagram (CMD, Figure \ref{fig:CMD}), additional checks were performed by eye to ensure these galaxies were indeed dusty spirals and not misclassified galaxies. 

The 50MGC provides the distance measurements used for all galaxies in this work and all but four galaxy stellar masses. [KK2000]~03, [KK2000]~53, BTS~109, and UGC~07242 have distance measurements available in the 50MGC but lack mass and galaxy color measurements. We use published masses from \citep{Hoyer2023a} for these galaxies after confirming that the published masses are consistent with the 50MGC for the remaining galaxies from this source. The lack of galaxy-wide colors for these four galaxies is not critical, as these values were only collected to place the sample galaxies on a CMD. Instead, we use the nuclear $g-i$ colors for these galaxies when plotting Figure~\ref{fig:CMD} (star symbols).
 
The CMD in Figure~\ref{fig:CMD} shows how our sample compares to the overall distribution of early- and late-type galaxies in the 50MGC (shaded contours). It also highlights the wide range in galaxy stellar mass spanned by our galaxies. For both early- and late-types, our galaxy masses extend down to $\lesssim10^{7}$ M$_\odot$ with good coverage over the lower-mass range ($7<$~log($M_{\rm gal}$)~$<10.5$). The nucleation fraction of our sample can also be visualized here via closed/open symbols (nucleated/not nucleated). All 23 late-types and the low-mass early-types are nucleated, while there is a mix in nucleation for high-mass early-types.

\subsection{Black Hole Masses} \label{sec:bh_masses}

Given that TDE rates are dominated \citep{StoneMetzger2016} by the density of stars within the MBH's sphere of influence (SOI; the region over which the MBH dominates the gravitational potential), we examined if our density measurements resolve this region in each galaxy. \newtext{We use a sphere of influence defined as the radius where the enclosed mass in stars equals the MBH mass \citep{Merritt2013}; note that this differs from the observational definition often used which depends on the velocity dispersion; these two are identical in the case of an isothermal sphere. Many of our sources lack velocity dispersion measurements, so we adopt the dispersion-independent SOI definition. For galaxies with velocity dispersion measurements, we find a median difference in influence radii of $\approx4.8$~pc between the two definitions.} 

When available, we use dynamical MBH mass estimates from \citet{Greene2020}, \citet{van-den-Bosch2016}, and \citet{Reines2015}, respectively (30 galaxies). Otherwise, we use MBH mass scaling relations with velocity dispersion (Eqs. \ref{eq:mbh_v_sig_g20_late} \& \ref{eq:mbh_v_sig_g20_early}) and host galaxy mass (Eqs. \ref{eq:mbh_v_mgal_g20_late} \& \ref{eq:mbh_v_mgal_g20_early}) from \citet{Greene2020} separated by type (61 galaxies). If there is an existing nuclear velocity dispersion measurement, we use the former (38 galaxies). It is worth noting that some velocity dispersion measurements are not directly comparable to the measurements used to construct the scaling relations (i.e. not measured within the galaxy's effective radius), but we use these for MBH mass estimates here as the velocity dispersion measurement has significantly lower scatter. Velocity dispersion data was collected from \citet{Greene2020}, \citet{Ho2009}, and HyperLEDA\footnote{http://leda.univ-lyon1.fr/}, respectively. 
The exact functional forms of the MBH scaling relations adopted from \citet{Greene2020} are given as:
\begin{equation} \label{eq:mbh_v_sig_g20_late}
    \rm{log}(\textit{M}_{\rm BH})_{\rm Late} = 7.4 + 2.54\times\rm{log}\left(\frac{\sigma}{160~[\rm{km/s}]}\right)
\end{equation}
\begin{equation} \label{eq:mbh_v_sig_g20_early}
    \rm{log}(\textit{M}_{\rm BH})_{\rm Early} = 8.03 + 4.24\times\rm{log}\left(\frac{\sigma}{160~[\rm{km/s}]}\right)
\end{equation}

\begin{equation} \label{eq:mbh_v_mgal_g20_late}
    \rm{log}(\textit{M}_{\rm BH})_{\rm Late} = 6.94 + 0.98\times\rm{log}\left(\frac{\textit{M}_{\rm{gal}}}{3\times10^{10}~[\rm{M}_{\odot}]}\right) 
\end{equation}
\begin{equation} \label{eq:mbh_v_mgal_g20_early}
    \rm{log}(\textit{M}_{\rm BH})_{\rm Early} = 7.89 + 1.33\times\rm{log}\left(\frac{\textit{M}_{\rm{gal}}}{3\times10^{10}~[\rm{M}_{\odot}]}\right)
\end{equation}

The corresponding SOI measurements are discussed further in Section~\ref{sec:soi}. NGC 5055 was excluded in this investigation based on an unreliable MBH mass measurement due to peculiar nuclear gas dynamics consistent with a bipolar outflow \citep{Blais-Ouellette2004}.

\section{Nuclear Mass-to-Light Ratios} \label{sec:ml}

Nuclear mass-to-light ($M/L$) ratios are necessary for converting SB measurements into mass densities. All $M/L$ ratios used in this work were derived in one of two ways. When available, we use nuclear $M/L$ ratios derived from SED modeling \citep[41 galaxies;][]{McDermid2015,Pechetti2017}. Otherwise, the color-$M/L$ relation from \citet{Taylor2011} was used to define $M/L$ ratios (50 galaxies). Published nuclear colors were used where available (30/50). For galaxies without existing nuclear color measurements (20/50), we measure them directly from archival \textit{HST} images (see Section \ref{sec:nuclear_colors} for details). All nuclear colors were transformed to SDSS g-i via Padova SSP models (details in Section \ref{sec:color_conversions}). See Table~\ref{tab:table_1} for an extensive list of bulk and nuclear properties for the sample galaxies, along with sources for the different measurements. 

\subsection{Nuclear Color Measurements} \label{sec:nuclear_colors}

For all galaxies lacking nuclear color information, we measure the colors from archival \textit{HST} images obtained from the Hubble Legacy Archive\footnote{Based on observations made with the NASA/ESA Hubble Space Telescope, and obtained from the Hubble Legacy Archive, which is a collaboration between the Space Telescope Science Institute (STScI/NASA), the Space Telescope European Coordinating Facility (ST-ECF/ESA) and the Canadian Astronomy Data Centre (CADC/NRC/CSA).} using aperture photometry. While only \newtext{20} galaxies require nuclear color definitions for $M/L$ ratio assignments, we measure colors for every galaxy lacking a prior measurement (61/91) for consistency. Two science images (level 3) were downloaded for each galaxy with one of the following camera and filter combinations: ACS/WFC $F475W$ \& $F850LP$, WFC3/UVIS $F475W$ \& $F814W$, or WFPC2/PC $F555W$ \& $F814W$. 

In each image, the galaxy's center was fine-tuned using a 2-D Gaussian fit via the ``centroid\_2dg" function of the ``photutils" Python library \citep{Bradley2021}. In cases where visual inspection revealed issues with the centroiding (13 galaxies), we aligned the galaxy centers using images in other filters. Once the galactic centers were well defined, we used the ``photutils.aperture\_photometry" function to extract the flux in an aperture with a radius of 0$\farcs$5. We performed aperture corrections to the resulting flux values using the recipes outlined in \citet{Gonzaga2010} for WFPC2 and the ACS \& WFC3 aperture correction tools \footnote{https://www.stsci.edu/\textit{HST}/instrumentation/acs/data-analysis/aperture-corrections} \footnote{https://www.stsci.edu/\textit{HST}/instrumentation/wfc3/data-analysis/photometric-calibration/uvis-encircled-energy}.

From this corrected flux, we convert to Vega magnitude using the zero-point for each camera/filter obtained from the ``acstools" python package for ACS, the STScI website\footnote{https://www.stsci.edu/\textit{HST}/instrumentation/wfc3/data-analysis/photometric-calibration/uvis-photometric-calibration} for WFC3, and the \textit{HST} Data Handbook for WFPC2. We also correct for extinction using reddening derived from \citet{Schlegel1998} dust maps with conversions to extinction for each camera/filter from \citet{Schlafly2011} for R$_\textrm{v}$ = 3.1. For each camera, our measured magnitudes agree with SExtractor aperture photometry within a few hundredths of a magnitude. 

Aperture photometry was impossible for 8 of the 61 galaxies lacking previously measured colors due to saturated images. Images were flagged as saturated if zero-valued pixels existed in the context frame within $0\farcs5$ of the galaxy center. For these galaxies, $g-i$ nuclear colors were derived using SDSS psfMags. This choice ensures the colors reflect each galaxy's nucleus, as the psfMags give the magnitude for the luminosity contained within the PSF. All non-saturated {\it HST} images utilized in this work can be found in MAST: \dataset[10.17909/d5ka-7566]{http://dx.doi.org/10.17909/d5ka-7566}.

Figure~\ref{fig:gal_v_nuc_gi} shows a comparison of nuclear $g-i$ colors vs galaxy-wide $g-i$ colors for our sample. The four galaxies lacking galaxy colors from 50MGC (see Section~\ref{sec:host_properties}) were excluded here. Clearly, the majority of our galaxies have redder nuclei when compared to the host galaxy. This can be interpreted in a few ways. First, it could indicate increased nuclear metallicity if one assumes similar ages for the nuclear and galactic stellar populations. This is not a poor assumption for high-mass early-type galaxies, and this increased nuclear metallicity has been observed in these galaxies (e.g. \citet{Neumayer2020,Fahrion2021}). On the other hand, this could indicate the presence of nuclear dust or a significantly older stellar population in the nucleus when compared to the overall galaxy.

\subsection{Color Conversions} \label{sec:color_conversions}
To use the color-$M/L$ relation and provide a consistent set of nuclear colors, all colors were converted to SDSS $g-i$. We make use of integrated magnitudes from a large grid of Padova Simple Stellar Population (SSP) models with ages spanning from $\sim$4~Myr to $\sim$25~Gyr and metallicities ranging from -2.0 to 0.3 to perform these conversions. The SSP models use stellar evolutionary tracks from PARSEC \citep{Bressan2012}, the YBC spectral library of \citet{Chen2019} with an updated spectrum for VEGA from \citet{Bohlin2020}, and a \citet{Kroupa2001} initial mass function. 
We use the magnitudes from these models under various photometric systems to construct color-color plots for each conversion. To ensure adequate sampling of the color-color space, we apply varying levels of extinction to the models with $0.0<E(B-V)<2.0$. A linear fit is performed to the color-color data over a range of 0.4 dex centered on the color to be converted. We then use the fit parameters to define the new color.

\section{Measuring Nuclear Densities} \label{sec:nuc_dens}

\subsection{\newtext{Deriving 3-D Density Profiles}}
We perform an Abel inversion of the SB profiles $I(R)$ along with the mass-to-light ratio ($\Upsilon$) to measure the 3-D stellar mass densities $\rho(r)$. Mathematically, this is described by
\begin{equation}
    \rho(r) = -\frac{1}{\pi}\int_{r}^{\infty}\frac{\textrm{d}I(R)}{\textrm{d}R}\frac{\textrm{d}R}{\sqrt{R^2-r^2}},
\end{equation}
where $I(R)$ is the 1-D SB profile as a function of projected radius $R$. 
\newtext{To ensure the shape of the SB profile is recovered accurately on all scales including both the nuclear star cluster and central regions of the galaxy,  we employ Multi-Gaussian Expansion (MGE) fits to parameterize the SB profiles via the $\mbox{\textsc{MgeFit}}$ Python package \citep{Cappellari2002-MGE}.}
\newtext{MGE models the input 1-D SB data as a linear combination of Gaussians, resulting in a flexible model capable of accurately fitting various profile shapes.} We use 1-D MGE fits via the ``mge\_fit\_1d" method, which provides the area ($A$) and width ($\sigma_{\rm g}$) of each Gaussian used to reconstruct the SB profile. 
These 1-D Gaussians are then converted to 2-D where the total luminosity is given as $L_{\textrm{tot}}=\sqrt{2\pi}A\sigma_{\rm g}$. The functional form of the resulting 2-D Gaussian SB profile is
\begin{equation} \label{eq:2dsb}
    I(R) = \frac{L_{\rm{tot}}}{2\pi\sigma_{\rm g}^2} e^{-R^2/{2\sigma_{\rm g}^2}}.
\end{equation}
Abel inversion of Equation~\ref{eq:2dsb} gives the 3-D mass profile for each Gaussian, as a function of spherical coordinate radius $r$, as
\begin{equation} \label{eq:rho}
    \rho(r) = \frac{\Upsilon L_{\rm{tot}}}{(\sqrt{2\pi}\sigma_{\rm g})^3}e^{-r^2/{2\sigma_{\rm g}^2}}.
\end{equation}
The density profiles defined by each component of the MGE fit are summed together to provide the final radial stellar mass density profile.

\subsection{\newtext{Accounting for Ellipticity}}

\newtext{The deprojection equations above assume spherical symmetry for the system.  
However, galaxy centers and resolved NSCs are not generally spherical systems, and this assumption will lead to an overestimate of the spherically averaged density; this spherically averaged density is the primary ingredient needed in our planned TDE rate modeling \citep[e.g.][]{StoneMetzger2016}.  
Therefore, we convert the radial density profile defined in Equation~\ref{eq:rho} to a spherically averaged value for an oblate ellipsoid. This is accomplished by first adjusting the total luminosity contained in the 2-D Gaussians. This entails multiplying $L_{\rm{tot}}$ by the observed axial ratio ($q'=b/a$, where $a$ and $b$ are the major and minor axes). To define the intrinsic axial ratios ($q$), we use a constant inclination of $60^\circ$, which is the median inclination for a random distribution of orientations:}  
\begin{equation}
    q^2 = \frac{q'^2 - {\rm sin}(60^\circ)^2}{{\rm cos}(60^\circ)^2}.
\end{equation}
\newtext{For an oblate ellipsoid defined with an intrinsic axial ratio $q$, Equation~\ref{eq:rho} then becomes:}
\begin{equation} \label{eq:rho_ell}
    \rho(r) = \frac{\Upsilon L_{\rm{tot}}q'}{(\sqrt{2\pi}\sigma_{\rm g})^3q}e^\frac{-r^2({\rm sin}(\phi)^2+{\rm cos}(\phi)^2/q^2)}{2\sigma_{\rm g}^2},
\end{equation}
\newtext{where $\phi$ is the polar angle.  Rather than integrating, we can instead use the symmetries in the system to evaluate this equation at $\phi=\pi/6$, which gives the spherically averaged density.}

\newtext{Because we are focused on the central density profiles of the galaxy, we use the central axial ratios for calculating the spherically averaged profiles.  We use a single value for each galaxy and draw this value from the innermost\footnote{Two galaxies from \citet{Pechetti2017}, NGC~3412 and NGC~4377, have ellipticity measurements incompatible with an inclination of $60^\circ$, so the inclination was adjusted to the lowest allowed values: $72^\circ$ and $71^\circ$, respectively.} measurements from the original sources of the profiles or profile fits in all cases\footnote{Note that for NGC~4486B, no value is given in \citet{Lauer2005}, thus we measured this from the HST images and found  $q'=0.55$ for the inner region of the galaxy}. 
Axial ratios for each galaxy are listed in Table~\ref{tab:table_1}; the median axial ratio is 0.9.  The resulting spherically averaged ellipsoidal density profiles differ from the purely spherical density profiles by a median of 13\% at 5~pc, with a maximum difference of 79\%. Therefore, this correction is typically small but is important since not correcting for this would lead to a systematic overestimate of the central densities in the more flattened systems.}


\subsection{Robustness of MGE Fits}
A Gaussian SB profile flattens out at small radii (i.e. its derivative approaches zero). This means we may be unable to accurately parameterize the innermost portions of our observed SB profiles using MGEs, as constant-density cores are unlikely to exist in the Kepler potential of the MBH\footnote{Specifically, a constant-density core can only exist in a Kepler potential if it has a large tangential anisotropy, which will generally relax away over a small fraction of a Hubble time \citep{Lezhnin2015}.}.  We, therefore, performed tests to gauge if a range of power-law profiles could be accurately recovered using MGE fits or, instead, if they were systematically flatter than the true profiles.  

We first defined various power-law SB profiles with slopes in the range [-3,-0.2] (corresponding to density slopes of [-4,-1.2]), which reflects the distribution of SB power-law slopes in our sample (measured at $r \lesssim 10$~pc). We then simulated what these profiles would look like using the radial sampling of each observed galaxy in our sample. For each power-law slope, we used $\mbox{\textsc{MgeFit}}$ to fit each simulated galaxy and then fit a power-law to each resulting MGE model (identical to the process we outline below for fitting our data). We varied the ``inner\_slope" and ``outer\_slope" parameters of $\mbox{\textsc{MgeFit}}$ and repeated this process. These parameters allow the user to force the slope of the MGE fit to a specific value at the smallest and largest radii, respectively. These comparisons indicate that values of 4 and 1 for “inner\_slope” and “outer\_slope,” respectively, produce the most accurate slope measurements across the range of slopes tested. Thus, we adopt these values for all MGE fits performed in this work. 

\begin{figure}[t]
    \centering
    \includegraphics[width=0.48\textwidth]{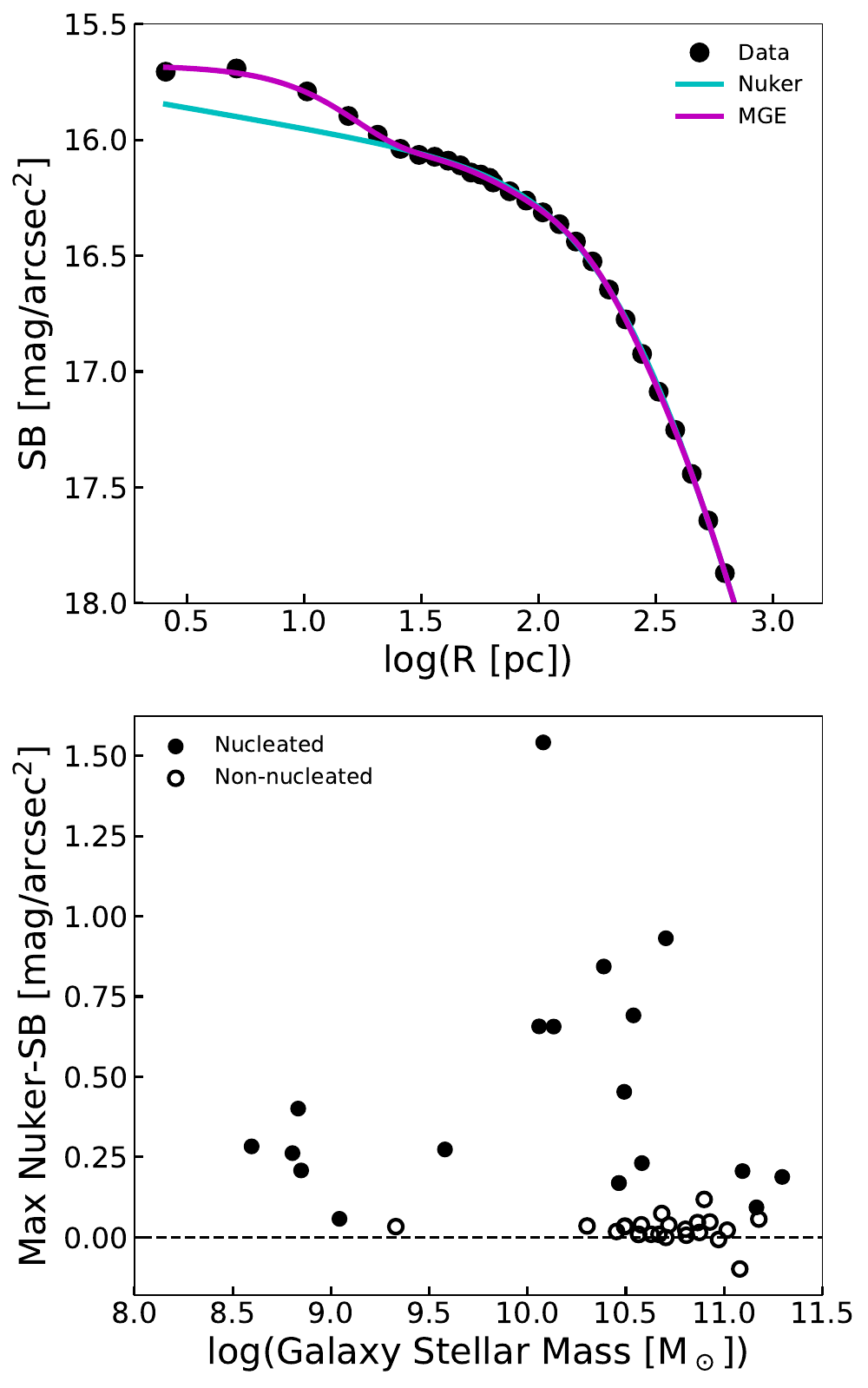} 
    \caption{Top: The MGE fit (magenta) to SB data for NGC~4365 (black points) compared to the prior Nuker-law fit (cyan) from \citet{Lauer2005} highlights the ability of an MGE to capture the excess nuclear luminosity resulting from the NSC. Middle: Maximum residual within a radius of 10~pc for Nuker-law fits plotted against galaxy stellar mass. The black-dashed line represents the $y=0$ line. This demonstrates the effect of nuclear star clusters on the Nuker-law fit quality at projected radii $< 10$~pc, where nucleated galaxies show the largest fit residuals. The corresponding MGE fits for these galaxies produce residuals $< 0.09$~mag/arcsec$^2$ in all cases.}
    \label{fig:nuke_v_mge}
\end{figure}

The ``ngauss" parameter of $\mbox{\textsc{MgeFit}}$ is also important, as it controls the number of Gaussians used when fitting. We also explored varying this parameter and found that significant differences between the true and MGE-derived slopes only occur for small values of ``ngauss," where the fit quality is compromised in general. We, therefore, utilize a large value of ``ngauss" (20) to ensure an accurate fit. Larger values of ``ngauss" were tested and found to show little improvement in fit statistics and slope recovery (``ngauss" = 50 only improved slope recovery by $10^{-4.7}$ when compared to ``ngauss" = 20). Using these $\mbox{\textsc{MgeFit}}$ parameters, the power-law profile slopes in our tests are all recovered to within 0.008 of the inserted slope value.  
Overall, this shows that our method of modeling galaxy SB profiles with MGEs does not significantly bias the recovered inner density profiles.

\subsubsection{Comparisons with Nuker-Law Parameterizations}
Many of the galaxies in our sample have previous “Nuker” law parameterizations for their SB profiles published in \citet{Lauer2007}, which are \newtext{designed to capture SB structure at larger radii}. Both \citet{Wang2004} and \citet{StoneMetzger2016} used these parameterizations to derive 3-D stellar densities for TDE rate calculations. However, a weakness of this approach is that the “Nuker” laws in \citet{Lauer2007} \newtext{did not, by design, include SB contributions from the NSC when present.}

The top panel of Figure \ref{fig:nuke_v_mge} shows the SB data (black) for one nucleated galaxy (NGC~1426) with a previous Nuker law fit (magenta). Our MGE fit to this data is shown in cyan, demonstrating that an MGE models the data at small radii much more accurately. This point is demonstrated further in the bottom panel, where the maximum residuals ($R < 10$~pc) for Nuker-law fits are plotted against galaxy stellar mass for galaxies drawn from the \citet{Lauer1995} and \citet{Lauer2005} samples. The Nuker-law residuals are clearly elevated in the nucleated galaxies, highlighting the model disagreement at small radii. Residuals in this radial range are significantly lower for the MGE fits ($<0.09$ mag/arcsec$^2$), with the largest improvements over Nuker-law fits observed in nucleated galaxies. This demonstrates that our use of MGE to parameterize the SB profiles can \newtext{(removed ``better" here)} recover density contributions from the NSC, and will therefore provide more accurate nuclear stellar densities for future TDE rate estimations (see Section \ref{sec:con-future}).

\section{Nuclear Density Scaling Relations} \label{sec:relations}

Here we describe how we derive and parameterize nuclear density profile scaling relations from our sample galaxies.  These relations are derived primarily with the goal in mind of calculating TDE rates.  We first discuss the role of NSCs in our sample galaxies and define the samples of galaxies over which we fit our nuclear density relations. We then describe our fitting method for the early- and late-type galaxy relations and present our best-fit relations, including their derived scatter.

\begin{figure}[t]
    \centering
    \includegraphics[width=0.5\textwidth]{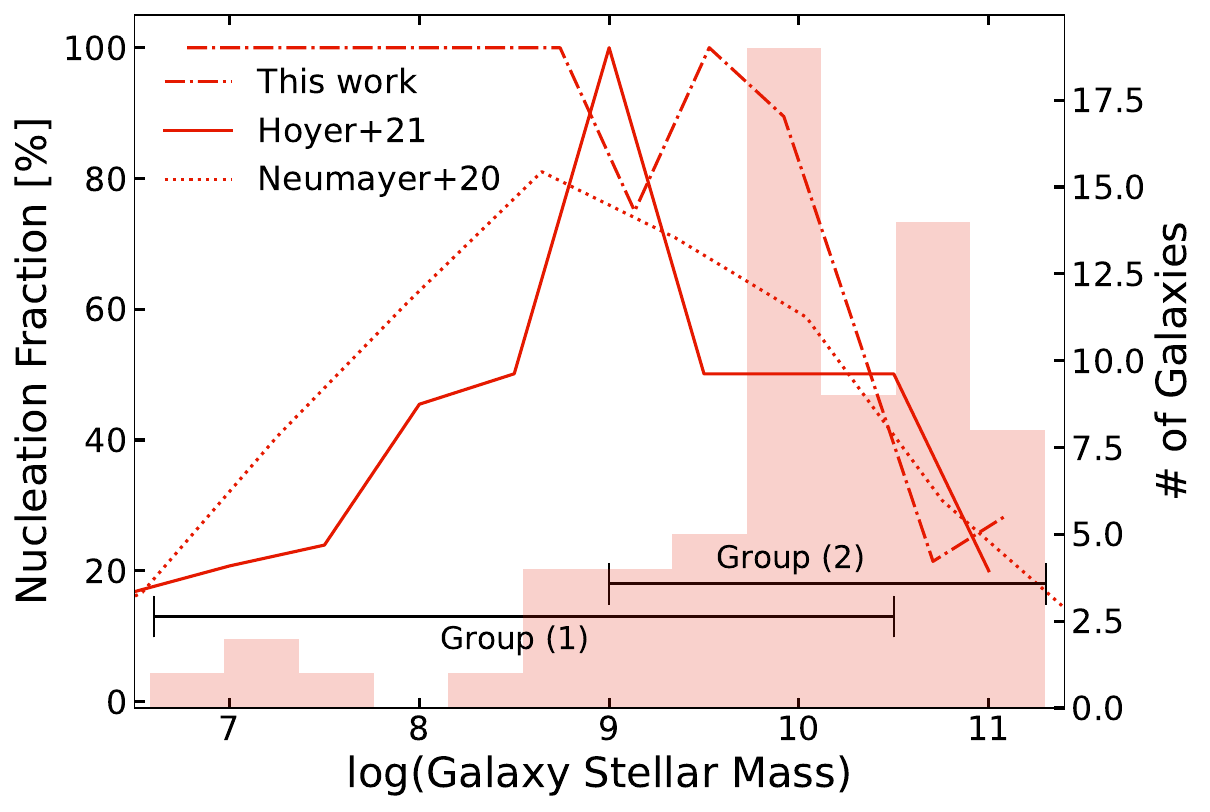} 
    \caption{This figure shows the nucleation fraction of early-type galaxies in our sample (dot-dash line) compared with existing measurements from \citet{Hoyer2021} and \citet{Neumayer2020} (solid and dotted, respectively). The distribution of early-type galaxies in our sample is shown by the histogram, and the black horizontal lines indicate the mass ranges used for the nucleated low-mass sample (Group 1) and the high-mass sample with mixed nucleation (Group 2). Over the mass range covered by Group 2, our sample follows a similar nucleation fraction to past observations.}
    \label{fig:nucleation}
\end{figure}

\subsection{Sample Division} \label{sec:division}
NSCs dominate the rates of TDEs in nearly all galaxies \citep{Pfister2020,Polkas2023}, and their influence increases as galaxy mass decreases. For low-mass galaxies, the TDE rates from the NSCs can be orders of magnitude higher than the rates from the host galaxies \newtext{alone}; therefore, it is reasonable to assume that only nucleated galaxies will contribute significantly to the volumetric TDE rate below a certain galaxy mass. Low-mass galaxies also greatly outnumber their high-mass counterparts, which further highlights their role in setting the volumetric TDE rate. Therefore, when using model galaxy samples to estimate volumetric TDE rates, accurately defining the stellar densities in nucleated low-mass galaxies is crucial.

To build the needed density relations for nucleated low-mass ($M_{\rm gal} < 10^{10.5}$~M$_\odot$) galaxies, we first divided the sample by morphology (early- and late-types). The sample was then divided further based on galaxy mass and nucleation. While many of the sources focused exclusively on nucleated galaxies, we also have sources that include both galaxies with and without NSCs. For sources that didn't focus solely on nucleated galaxies \citep{Lauer1995,Lauer2005,Pechetti2017}, missing nucleation information was first collected from previous literature \citet{Cote2006,Georgiev2014,Sanchez-Janssen2019}. Seven galaxies from \citet{Pechetti2017} lacked prior NSC designations: NGC~3412, NGC~3522, NGC~3796, NGC~4342, NGC~4733, PGC~028887, and PGC~050395. For these galaxies, nucleation was determined by eye from the SB data, where a significant increase in brightness above the inward extrapolation of the underlying galaxy light profile indicates the presence of an NSC. We found all galaxies other than PGC~028887 had an NSC.  The division of the sample into nucleated (filled circles) and non-nucleated galaxies (open circles) is shown in Figure~\ref{fig:CMD}; this same convention is used throughout the paper.

All late-type galaxies in our sample host NSCs and have masses $< 10^{10.5}$~M$_\odot$, except for one (NGC~7727). Therefore, we exclude this galaxy from the relations and define just one subsample for late-types satisfying our low-mass nucleated galaxies requirement (22 galaxies). The early-type galaxies in our sample consist of both nucleated and non-nucleated galaxies with many galaxy masses $> 10^{10.5}$~M$_\odot$. Thus, we divide the early-types into two groups: (1) nucleated low-mass ($< 10^{10.5}$~M$_\odot$; 44 galaxies) and (2) mixed nucleation higher-mass ($10^9$~M$_\odot$ $<$ $M_{\rm gal}$ $< 10^{11.3}$~M$_\odot$; 59 galaxies). Group (1) was defined as an early-type subsample consistent with the late-type galaxy sample, which together, produce the crucial nuclear density relations for low-mass nucleated galaxies. 

Group (2), on the other hand, was defined to provide density relations for high-mass early-type galaxies. When considering the volumetric TDE rate, high-mass galaxies are unlikely to dominate the overall TDE host population \citep[e.g.][]{StoneMetzger2016}. Not only are they outnumbered by low-mass galaxies, but two other factors work against them.  First, high-mass galaxies generally have lower central densities and longer relaxation times, reducing TDE rates \citep{Wang2004, StoneMetzger2016}.  Second, high-mass galaxies generally host more massive MBHs resulting in TDE rate suppression due to direct capture (i.e. the TDE occurs within the event horizon producing no observable flare; \citealt{Hills1975}). For Solar mass stars, the Hills mass is reached around non-spinning MBHs with masses $\geq 10^{8}$~M$_\odot$.  This effect has been observed in empirical TDE rate measurements \citep{vanVelzen2018, Yao2023}. However, near-future TDE samples numbering in the thousands \citep{Bricman2020, Ben-Ami2023} will probe even subdominant contributions of TDEs from high-mass galaxies, and it will therefore be important to understand TDE rates from such systems, which is why we estimate central density scaling relations for them as well.

When defining a scaling relation for a galaxy sample with mixed nucleation, it is important that the nucleation fraction reflects observations to avoid biases. Thus, we compare the nucleation fraction of early-type galaxies in our sample to previous measurements from \citet{Hoyer2021} and \citet{Neumayer2020} in Figure \ref{fig:nucleation}. The black horizontal lines show the mass ranges of Groups (1) and (2). Over the mass range covered by Group (2), our nucleation fraction decays at high masses, similar to observations. We note that our sample shows higher nucleation for galaxies with masses $\sim10^{10}$~M$_\odot$. At low galaxy masses (Group 1), our nucleation fraction approaches 100\%, which is caused by our preference for nucleated low-mass galaxies during sample construction. Given that the nucleation fraction of Group (2) roughly agrees with observations, the relations defined by this subsample will provide a nucleation-independent relation suited for high-mass early-type galaxies. We note that there is an overlap between Groups (1) and (2) due to the mass range defined for Group (2), which extends to the lowest mass where the nucleation fractions are consistent. While we do not have a corresponding relation for late-type galaxies, the nucleation fraction for high-mass late-types is not observed to decline at the highest masses \citep[e.g][]{Neumayer2020, Hoyer2021}. There are also fewer of these galaxies than high-mass early-types \citep[e.g][]{Baldry2012, Driver2022}. Therefore, the low-mass nucleated relation for late-types can be reasonably used for high-mass late-type galaxies as well.

\begin{figure*}[t]
    \centering
    \includegraphics[width=0.9\textwidth]{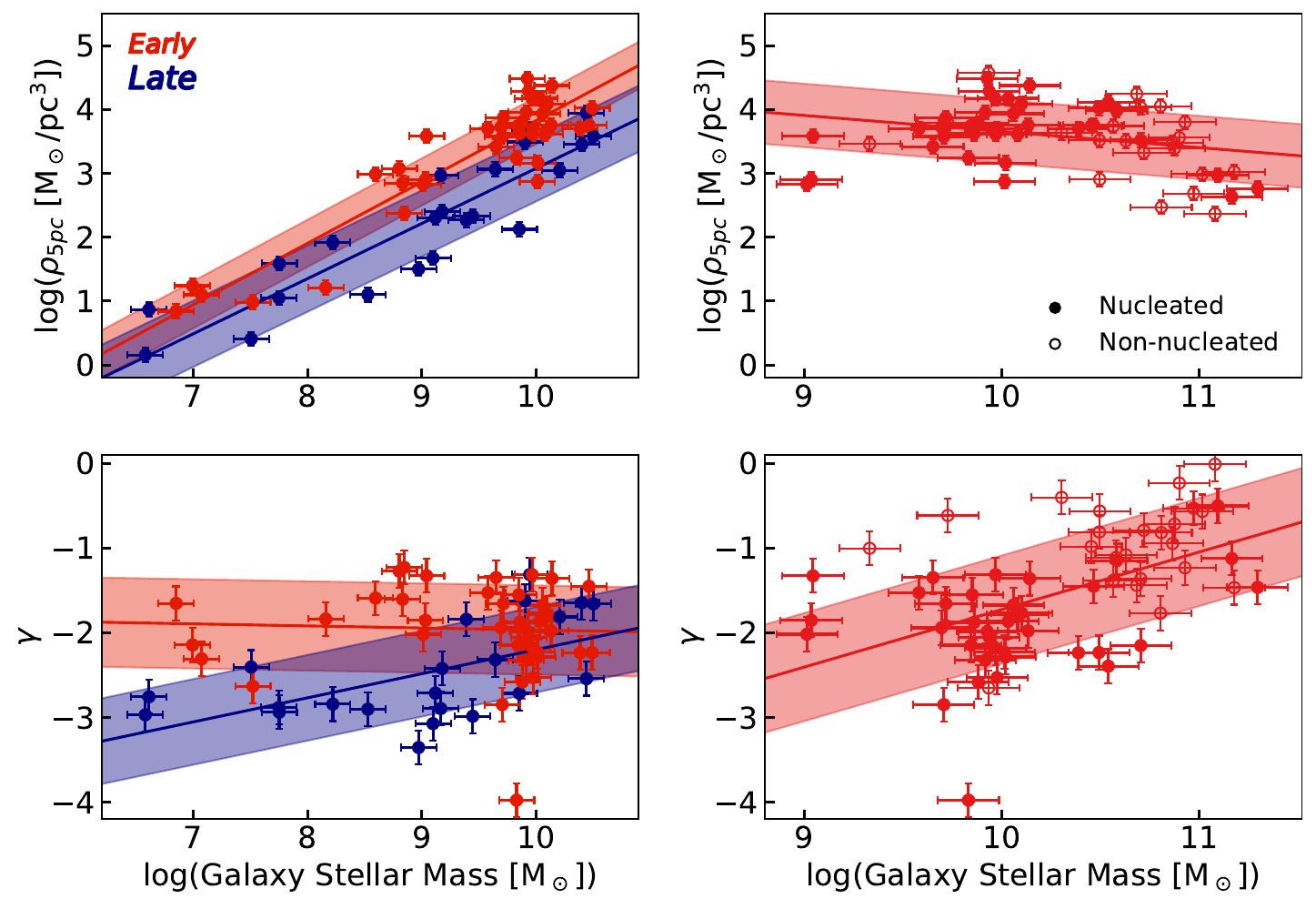} 
    \caption{New nuclear scaling relations derived in this work. The left two panels display the relations between $\rho_{\rm 5pc}$ (top) and $\gamma$ (bottom) and galaxy mass for the nucleated galaxy subsamples, where the colors distinguish galaxy type. The right panels are identical to the left but show the relations for the higher-mass early-type galaxies with mixed nucleation.}
    \label{fig:relations}
\end{figure*}

\subsection{Fitting the Relations}
To construct relations between the \newtext{nuclear} density structures and host galaxy properties, we parameterize the densities by fitting the inner regions with a simple power-law. The primary parameters are the density at 5pc ($\rho_{\rm 5pc}$) and the power-law slope ($\gamma$) over the inner region of the galaxy/NSC. Specifically, the slopes are fit over a minimum physical radius of 10 pc if this is greater than/equal to the angular radius of the third SB data point (80\% of cases). Otherwise, the fits are performed over a fixed angular scale defined as twice the \textit{HST} pixel scale, ensuring that 3 original SB measurements are included. For the 18 galaxies using fixed angular scales for slope measurements, the median angular scale corresponds to $\sim$12~pc with a maximum of 22~pc. All fits were visually inspected to confirm fit quality and ensure the fits were not averaging over abrupt variations in the SB profiles. We use the $\mbox{\textsc{linmix}}$ Python package to fit the linear relations between $\rho_{\rm 5pc}$ \& $\gamma$ and host galaxy stellar mass, which also models the scatter in the relations. The functional forms of these fits are given in Section~\ref{sec:new_relations}.

For uncertainties on $\rho_{\rm 5pc}$, we adopt the median error estimate from similar $\rho_{\rm 5pc}$ measurements in \citet{Pechetti2020} (0.11), which is dominated by uncertainties in the distances and color-based $M/L$ ratios. 
\newtext{To gauge the uncertainties on $\gamma$, we used a combination of independent SB measurements of our sample galaxies, and varying the fit range for galaxies with more than 3 SB measurements within 10 pc. While 4 galaxies have multiple SB measurements \citep[from][]{Lauer2005, Pechetti2020}, NGC 7457 was excluded here due to PSF deconvolution issues in the SB profile from \citet{Lauer2005}. The 3 remaining independent SB measurements give variations in $\gamma$ of $\approx0.2$. For the 11 galaxies with $>3$ SB measurements, we varied the maximum fit radius from the 3rd SB data point (our minimum fitting requirement) out to 10~pc. This investigation produced variations in $\gamma$ that were generally smaller than those found from the more robust independent SB measurements. Although rare, varying the fit range did produce differences in $\gamma$ as high as $\approx0.3$. Based on these investigations, we adopt an error of 0.2 for all $\gamma$ measurements but caution that this is approximate.}
Additionally, we assume a general 30\% uncertainty on the galaxy stellar masses. We note that doubling this error results in no significant difference in the final fit parameters or scatter.

\subsection{The New Relations} \label{sec:new_relations}

The fits for the new density scaling relations are shown in Figure~\ref{fig:relations} with the 1-$\sigma$ scatter indicated by the shaded regions. From the top left panel, it is clear that \newtext{nuclear} densities for both early- and late-type \newtext{nucleated galaxies} are  correlated with host galaxy stellar mass, with slightly higher densities in early-types.
At higher stellar masses ($\sim$10$^{10}$~M$_\odot$), \newtext{nuclei} in early-type \newtext{NSC-host} galaxies have nearly an order of magnitude higher 3-D densities than late-types, but at lower galaxy masses, this difference is much smaller. \newtext{The density slope relations for nucleated galaxies} (lower left panel) show significantly more scatter than the densities and a stronger dependence on galaxy type. For late-type galaxies, $\gamma$ values are typically lower when compared to early-types, with the difference becoming more pronounced in low-mass galaxies.  \newtext{Nucleated} early-types have typical density slopes $\gamma \sim -2$ across all galaxy stellar masses. 

The upper right panel of Figure~\ref{fig:relations} reveals an opposite trend for nuclear densities of the higher mass early-type galaxies with mixed nucleation. There is also a higher degree of scatter in this relation when compared to the low-mass nucleated galaxy sample. The density slope relation for the high-mass early-types (lower right panel) shows comparable scatter to the nucleated-only sample along with a similar, but steeper, trend with galaxy mass.

The functional forms of all nuclear density scaling relations fit in this work are given below. \\
\\
\noindent \textbf{Nucleated relations:}
\begin{equation} \label{eq:et_nsc_dens}
    \rm{log}(\rho_{5pc})_{\rm{Early}} = 0.96\pm{0.07}\times\rm{log}\left(\frac{\textit{M}_{\rm{gal}}}{10^{9}~[\rm{M}_{\odot}]}\right) + 2.86\pm{0.07}
\end{equation}
\begin{equation} \label{eq:lt_nsc_dens}
    \rm{log}(\rho_{5pc})_{\rm{Late}} = 0.86\pm{0.10}\times\rm{log}\left(\frac{\textit{M}_{\rm{gal}}}{10^{9}~[\rm{M}_{\odot}]}\right) + 2.21\pm{0.12}
\end{equation}
\begin{equation} \label{eq:et_nsc_slope}
    \gamma_{\rm{~Early}} = 0.03\pm{0.10}\times\rm{log}\left(\frac{\textit{M}_{\rm{gal}}}{10^{9}~[\rm{M}_{\odot}]}\right) - 1.94\pm{0.10}
\end{equation}
\begin{equation} \label{eq:lt_nsc_slope}
    \gamma_{\rm{~Late}} = 0.28\pm{0.10}\times\rm{log}\left(\frac{\textit{M}_{\rm{gal}}}{10^{9}~[\rm{M}_{\odot}]}\right) - 2.49\pm{0.12}.
\end{equation}
\\
\noindent \textbf{Mixed-nucleation relations:}
\begin{equation} \label{eq:eth_dens}
    \rm{log}(\rho_{5pc})_{\rm{Early}} = -0.25\pm{0.13}\times\rm{log}\left(\frac{\textit{M}_{\rm{gal}}}{10^{9}~[\rm{M}_{\odot}]}\right) + 3.45\pm{0.10}
\end{equation}
\begin{equation} \label{eq:eth_slope}
    \gamma_{\rm{~Early}} = 0.68\pm{0.17}\times\rm{log}\left(\frac{\textit{M}_{\rm{gal}}}{10^{9}~[\rm{M}_{\odot}]}\right) - 1.18\pm{0.13}.
\end{equation}

The intrinsic scatters for these relations are 0.37, 0.52, 0.53, 0.51, 0.50, and 0.64, respectively. 

\begin{figure}[t] 
    \centering
    \includegraphics[width=0.5\textwidth]{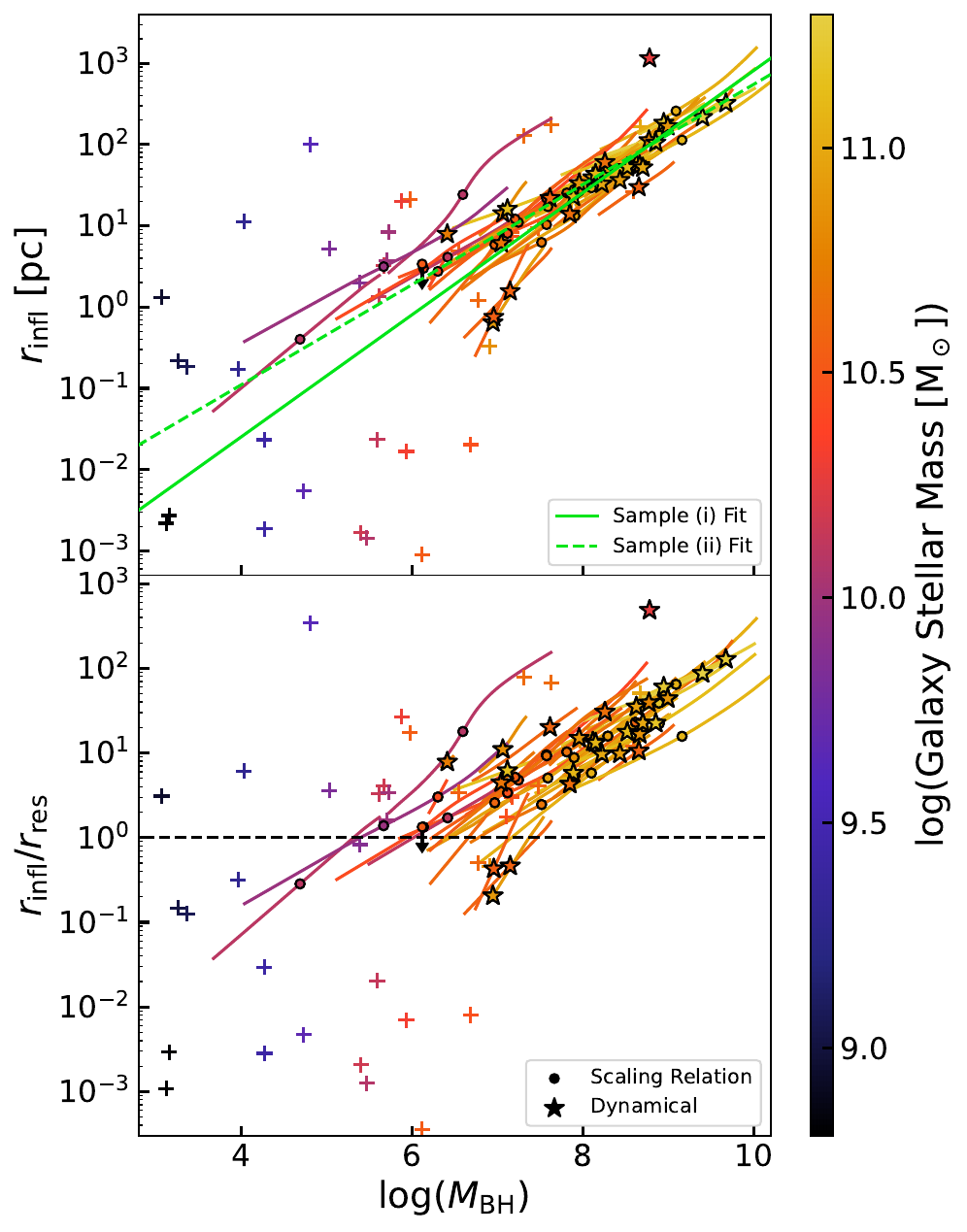} 
    \caption{Top: Influence radii for all sample galaxies plotted against MBH mass and color-coded by the galaxy stellar mass. Star symbols indicate galaxies with dynamical MBH mass measurements, while bullets \newtext{and crosses} indicate MBH mass estimates from scaling relations. The diagonal lines show the range of $r_{\rm infl}$ possible given uncertainties in these mass estimates (either dynamical or from scaling relations). 
    \newtext{The galaxies plotted as crosses have distributions of $r_{\rm infl}$ spanning $\gtrsim2$ orders of magnitude, so they are plotted without a line to avoid clutter.}
    NGC 4733 is plotted as an upper limit due to the use of a constant density extrapolation. The solid and dashed green lines provide the two power-law fits to the data points. Bottom: Identical to the top panel but displays the fraction of the influence radius that is observationally resolved, with the black dashed line indicating the threshold for influence radius resolution.}
    \label{fig:r_influence}
\end{figure}

\section{Do our Density Measurements Probe the MBH Sphere of Influence?} \label{sec:soi}

In the future, we will use these new empirical relations in tandem with existing galaxy demographic relations to simulate the nuclear density profiles of a realistic volume-limited galaxy sample that can be used for TDE rate modeling. As mentioned previously, TDE rates depend heavily on the density of stars near the MBH’s SOI \citep{StoneMetzger2016}. We therefore test how well our SB data resolve the MBH SOI ($r_{\rm infl}$) in our sample galaxies. We first use our MBH mass estimates (see Section~\ref{sec:bh_masses}) to determine $r_{\rm infl}$ in each galaxy. Then, we combine this with the minimum resolved radius ($r_{\rm res}$) to determine whether or not the SOI was resolved.

To compute the influence radii, the enclosed mass needs to be defined as a function of radius for each galaxy and requires inward extrapolation of the density profiles beyond our resolution limits. For each galaxy, we extrapolate our power-law density fits to a minimum radius of $10^{-3}$~pc. One galaxy, NGC 4733, has such a steep density profile (-3.96) that extrapolation to this scale results in $r_{\rm infl}$ values less than our minimum extrapolated radius. In general, it is unphysical for nuclear density profiles to be characterized by power-laws with very large $\gamma$ values over an arbitrary range of radii \citep{Stone2018}. When extrapolating to the center of the galaxy, if $\gamma \ge 2$, diffusion of stars through energy space becomes strongly non-local; if $\gamma \ge 9/4$, the TDE rate actually diverges as one considers more tightly bound stars; if $\gamma \ge 3$, the stellar mass enclosed diverges. For all of these ``ultra-steep'' values of $\gamma$, relaxation times are shortest at very small radii, and theoretically one expects the system to relax into a quasi-steady state from the inside out, achieving a broken power-law density profile on small scales \citep{Stone2018}, with the inner power-law slope for stars resembling the classic Bahcall-Wolf solution ($\gamma \approx 1.5-1.75$; \citealt{Bahcall1976, Bahcall1977}). Thus, to set an upper limit on the influence radius for NGC 4733, we employ a constant density extrapolation for this galaxy. 

Figure~\ref{fig:r_influence} displays the results of this investigation.
The data points in both panels display the MBH mass for each galaxy, along with their corresponding influence radius.
The bottom panel shows the fraction of $r_{\rm infl}$ resolved by our observations (Section~\ref{sec:bh_masses}). The diagonal lines on some points translate the error on the dynamical BH mass measurements into the log($M_{\rm BH}$)- $r_{\rm infl}$ plane.  For galaxies with MBH masses from scaling relations, we use the scatter in the scaling relations to define these lines. \newtext{Galaxies plotted as crosses have influence radii distributions spanning $\gtrsim2$ orders of magnitude based on the scatter in the scaling relation, and thus are plotted as single points to aid readability.}
The two green lines in the upper panel of the figure show the power-law fits to the data, which are discussed further in Section \ref{sec:mbh_v_rinf}.

From the bottom panel of Figure~\ref{fig:r_influence}, it is clear that the MBH SOI is resolved by our density measurements in a majority of galaxies, especially at the high-mass end. At MBH masses above $\sim10^6$~M$_\odot$, almost all of our galaxies are well resolved, while the picture is less clear at lower masses. For those with weaker constraints, the scatter in the relations used to define the MBH mass distributions provides cases where the SOI may have been resolved.

\subsection{$M_{\rm BH}$ vs. $r_{\rm infl}$ Relation} \label{sec:mbh_v_rinf}
To aid theorists and modelers, an empirical power-law relationship between MBH mass and influence radius was first introduced in the appendix of \citet{StoneMetzger2016}. For comparison, we fit an identical power-law relation to our data. To gauge uncertainties in these fit parameters, we perform 1000 rounds of bootstrap resampling and refit each time. The fits were performed on two subsamples: (i) galaxies with dynamical MBH mass measurements (30 galaxies; $9.3 <$ M$_{gal} < 11.3$) and (ii) galaxies with dynamical or $M_{\rm BH}-\sigma$ defined MBH masses (68 galaxies; $9.0 <$ M$_{gal} < 11.3$). The resulting relations follow the parametric form below:

\begin{equation}
    \rm{log}(\textit{r}_{\rm infl}[pc]) = \textit{A}~ \rm{log}\left(\frac{\textit{M}_{BH}}{10^8 [M_\odot]}\right) + \textit{B}.
\end{equation}

These fits are shown as green lines in Figure \ref{fig:r_influence}. For the dynamical mass subsample (i), we get $A = 0.75 \pm 0.12$ and $B = 1.40 \pm 0.08$, while the ``full" sample (ii) gives $A = 0.62 \pm 0.10$ and $B = 1.51 \pm 0.05$. \newtext{The fitted slope values ($A$) from both of our subsamples agree with the full galaxy sample relations presented in \citet{StoneMetzger2016} within error ($A_{\rm SM16} = 0.69$), while the intercepts ($B$) for our relations are slightly higher ($B_{\rm SM16} = 1.20$).}

\section{Conclusions \& Future Work} \label{sec:con-future}

We have assembled the largest dataset of resolved galactic nuclei where the 3-D stellar mass densities can be characterized on pc-scales (91 galaxies). These measurements are based on both PSF-deconvolved SB profiles \citep{Lauer1995,Lauer2005} and model estimates that include PSF fitting \citep{Pechetti2017,Pechetti2020,Hoyer2023a}. These 3-D stellar densities were deprojected via an Abel inversion of MGE fits to the SB data paired with a measured $M/L$ ratio. We make use of stellar population synthesis derived $M/L$ ratios where available \citep{McDermid2015,Pechetti2017} and supplement missing measurements with nuclear color-based $M/L$ estimates. We then characterized the inner $\sim10$~pc of each density profile with a power-law fit, with the fit parameters being the 3-D density at $r=5$~pc ($\rho_{\rm 5pc}$) and the slope ($\gamma$).

To construct useful nuclear density scaling relations for future rate estimates of dynamics-driven nuclear transients, we disaggregate our sample by galaxy mass, nucleation, and galaxy type. We anticipate this will be most useful for estimating TDE rates from loss cone physics, but it may be applicable to other transients of dynamical origin, such as LIGO/Virgo-band gravitational wave sources \citep{Antonini2016}, extreme mass-ratio inspirals \citep{Hopman2005, Qunbar2023}, and X-ray quasi-periodic eruptions \citep{Miniutti2019, Wevers2022}. Focusing on TDEs, it has been shown that in nucleated galaxies, these events are usually dominated by the stars of the NSC \citep[e.g.][]{Pfister2020,Polkas2023}. When considering the increasing number of galaxies at lower masses \citep[e.g.][]{Baldry2012,Driver2022} and the prevalence of NSCs in these galaxies \citep[e.g.][]{Sanchez-Janssen2019,Neumayer2020,Hoyer2021}, it is likely that most TDEs will occur in nucleated galaxies. We therefore define two subsamples of NSC host galaxies (early- and late-type) with stellar masses $< 10^{10.5}$~M$_\odot$). 

In higher-mass early-type galaxies, high nuclear densities suggest they may have significant rates of TDEs even when not nucleated. We therefore construct a third subsample for higher-mass ($10^9$~M$_\odot$ $<$ $M_{\rm gal}$ $< 10^{11.3}$~M$_\odot$ early-type galaxies with mixed nucleation aimed at producing relations to aid in constraining the subdominant population of TDEs expected from these galaxies. Although we do not have a comparable subsample for late-types, extrapolation of the nucleated-only relation for late-types is not unreasonable, given the prevalence of NSCs in the higher-mass spirals.

For each of the subsamples, we fitted scaling relations between the host galaxy stellar mass and the central density parameters ($\rho_{\rm 5pc}$ and $\gamma$), which are shown in Figure \ref{fig:relations}. The low-mass nucleated galaxies that likely dominate volumetric TDE rates show a positive relation between stellar mass densities at 5~pc and galaxy mass. At the highest masses, nucleated early-types have densities roughly an order of magnitude higher than the late-types, and these differences decrease at lower galaxy masses. There are also significant morphological differences in the nucleated stellar density slopes, with late-type galaxies featuring steeper density profiles, especially at lower masses. The high-mass early-type subsample with mixed nucleation produces an opposite trend in central densities and a steeper relation for the central density gradients.

Compared to past samples of galactic nuclei used for loss cone modeling \citep{StoneMetzger2016,Wang2004}, we find that employing an MGE on small scales greatly improves the nuclear region's quality of fit relative to parametric models that fit the SB profile on larger scales.  We therefore find higher densities than these previous works in nucleated galaxies.  

Because TDE rates depend primarily on the density of stars within the MBH's SOI, we also investigated if the SOI was resolved in each galaxy (assuming an MBH mass from dynamical measurements or scaling relations). We find that for MBH masses above $\sim10^6$~M$_\odot$, almost all of our galaxies have well-resolved influence radii. For lower mass MBHs, the resolution is less certain given the wide distribution of possible influence radii due to scatter in the MBH scaling relations. 

In \newtext{Papers II \& III} (Hannah et al., {\em in prep}), these relations will be combined with existing galaxy demographic relations to construct realistic model galaxy samples used to estimate the intrinsic volumetric TDE rate. Before modeling TDE rates, the galaxies in each sample will be populated with central MBHs based on present-day expectations from different theoretical MBH formation scenarios. TDE rates will computed for each model galaxy using our Python implementation of a standard steady-state loss-cone formalism \citet{Wang2004, StoneMetzger2016}, which models the diffusion of stars in angular momentum via 2-body interactions. The volumetric TDE rates derived from these models will then be forward-modeled into detection rates for the Zwicky Transient Facility (ZTF) and the Rubin Observatory (LSST) for comparisons with observed TDE rates. This will potentially place constraints on both the unknown MBH demographics in low-mass galaxies and even the formation mechanism of MBHs.

\section*{Acknowledgements}
C.H.H. and A.C.S. acknowledge support from National Science Foundation astronomy and astrophysics grant AST-2108180. N.C.S. acknowledges financial support from the Israel Science Foundation (Individual Research Grant 2565/19) and the Binational Science foundation (grant Nos. 2019772 and 2020397).  C.H.H. would also like to thank Todd Lauer, Renuka Pechetti, and Nils Hoyer for all their help with obtaining the necessary surface brightness measurements.

\bibliography{references}

\begin{thebibliography}{}
\expandafter\ifx\csname natexlab\endcsname\relax\def\natexlab#1{#1}\fi
\providecommand{\url}[1]{\href{#1}{#1}}
\providecommand{\dodoi}[1]{doi:~\href{http://doi.org/#1}{\nolinkurl{#1}}}
\providecommand{\doeprint}[1]{\href{http://ascl.net/#1}{\nolinkurl{http://ascl.net/#1}}}
\providecommand{\doarXiv}[1]{\href{https://arxiv.org/abs/#1}{\nolinkurl{https://arxiv.org/abs/#1}}}

\bibitem[{{Agarwal} {et~al.}(2012){Agarwal}, {Khochfar}, {Johnson}, {Neistein}, {Dalla Vecchia}, \& {Livio}}]{Agarwal2012}
{Agarwal}, B., {Khochfar}, S., {Johnson}, J.~L., {et~al.} 2012, \mnras, 425, 2854, \dodoi{10.1111/j.1365-2966.2012.21651.x}

\bibitem[{{Antonini} \& {Rasio}(2016)}]{Antonini2016}
{Antonini}, F., \& {Rasio}, F.~A. 2016, \apj, 831, 187, \dodoi{10.3847/0004-637X/831/2/187}

\bibitem[{{Bahcall} \& {Wolf}(1976)}]{Bahcall1976}
{Bahcall}, J.~N., \& {Wolf}, R.~A. 1976, \apj, 209, 214, \dodoi{10.1086/154711}

\bibitem[{{Bahcall} \& {Wolf}(1977)}]{Bahcall1977}
---. 1977, \apj, 216, 883, \dodoi{10.1086/155534}

\bibitem[{{Baldry} {et~al.}(2012){Baldry}, {Driver}, {Loveday}, {Taylor}, {Kelvin}, {Liske}, {Norberg}, {Robotham}, {Brough}, {Hopkins}, {Bamford}, {Peacock}, {Bland-Hawthorn}, {Conselice}, {Croom}, {Jones}, {Parkinson}, {Popescu}, {Prescott}, {Sharp}, \& {Tuffs}}]{Baldry2012}
{Baldry}, I.~K., {Driver}, S.~P., {Loveday}, J., {et~al.} 2012, \mnras, 421, 621, \dodoi{10.1111/j.1365-2966.2012.20340.x}

\bibitem[{{Begelman}(2008)}]{Begelman2008}
{Begelman}, M.~C. 2008, in American Institute of Physics Conference Series, Vol. 990, First Stars III, ed. B.~W. {O'Shea} \& A.~{Heger}, 489--493, \dodoi{10.1063/1.2905669}

\bibitem[{{Begelman} {et~al.}(2006){Begelman}, {Volonteri}, \& {Rees}}]{Begelman2006}
{Begelman}, M.~C., {Volonteri}, M., \& {Rees}, M.~J. 2006, \mnras, 370, 289, \dodoi{10.1111/j.1365-2966.2006.10467.x}

\bibitem[{{Bi} {et~al.}(2020){Bi}, {Feng}, \& {Ho}}]{Bi2020}
{Bi}, S., {Feng}, H., \& {Ho}, L.~C. 2020, \apj, 900, 124, \dodoi{10.3847/1538-4357/aba761}

\bibitem[{{Blais-Ouellette} {et~al.}(2004){Blais-Ouellette}, {Amram}, {Carignan}, \& {Swaters}}]{Blais-Ouellette2004}
{Blais-Ouellette}, S., {Amram}, P., {Carignan}, C., \& {Swaters}, R. 2004, \aap, 420, 147, \dodoi{10.1051/0004-6361:20034263}

\bibitem[{{Bohlin} {et~al.}(2020){Bohlin}, {Hubeny}, \& {Rauch}}]{Bohlin2020}
{Bohlin}, R.~C., {Hubeny}, I., \& {Rauch}, T. 2020, \aj, 160, 21, \dodoi{10.3847/1538-3881/ab94b4}

\bibitem[{Bradley {et~al.}(2021)Bradley, Sip{\H o}cz, Robitaille, Tollerud, Vin{\'\i}cius, Deil, Barbary, Wilson, Busko, Donath, G{\"u}nther, Cara, krachyon, Conseil, Bostroem, Droettboom, Bray, Lim, Bratholm, Barentsen, Craig, Rathi, Pascual, Perren, Georgiev, de~Val-Borro, Kerzendorf, Bach, Quint, \& Souchereau}]{Bradley2021}
Bradley, L., Sip{\H o}cz, B., Robitaille, T., {et~al.} 2021, astropy/photutils: 1.3.0, 1.3.0,  Zenodo, \dodoi{10.5281/zenodo.5796924}

\bibitem[{{Bressan} {et~al.}(2012){Bressan}, {Marigo}, {Girardi}, {Salasnich}, {Dal Cero}, {Rubele}, \& {Nanni}}]{Bressan2012}
{Bressan}, A., {Marigo}, P., {Girardi}, L., {et~al.} 2012, \mnras, 427, 127, \dodoi{10.1111/j.1365-2966.2012.21948.x}

\bibitem[{{Bricman} \& {Gomboc}(2020)}]{Bricman2020}
{Bricman}, K., \& {Gomboc}, A. 2020, \apj, 890, 73, \dodoi{10.3847/1538-4357/ab6989}

\bibitem[{{Bromm} \& {Larson}(2004)}]{Bromm2004}
{Bromm}, V., \& {Larson}, R.~B. 2004, \araa, 42, 79, \dodoi{10.1146/annurev.astro.42.053102.134034}

\bibitem[{{Bromm} \& {Loeb}(2003)}]{Bromm2003}
{Bromm}, V., \& {Loeb}, A. 2003, \apj, 596, 34, \dodoi{10.1086/377529}

\bibitem[{{Chen} {et~al.}(2019){Chen}, {Girardi}, {Fu}, {Bressan}, {Aringer}, {Dal Tio}, {Pastorelli}, {Marigo}, {Costa}, \& {Zhang}}]{Chen2019}
{Chen}, Y., {Girardi}, L., {Fu}, X., {et~al.} 2019, \aap, 632, A105, \dodoi{10.1051/0004-6361/201936612}

\bibitem[{{Cohn} \& {Kulsrud}(1978)}]{Cohn1978}
{Cohn}, H., \& {Kulsrud}, R.~M. 1978, \apj, 226, 1087, \dodoi{10.1086/156685}

\bibitem[{{C{\^o}t{\'e}} {et~al.}(2006){C{\^o}t{\'e}}, {Piatek}, {Ferrarese}, {Jord{\'a}n}, {Merritt}, {Peng}, {Ha{\c{s}}egan}, {Blakeslee}, {Mei}, {West}, {Milosavljevi{\'c}}, \& {Tonry}}]{Cote2006}
{C{\^o}t{\'e}}, P., {Piatek}, S., {Ferrarese}, L., {et~al.} 2006, \apjs, 165, 57, \dodoi{10.1086/504042}

\bibitem[{{Devecchi} \& {Volonteri}(2009)}]{Devecchi2009}
{Devecchi}, B., \& {Volonteri}, M. 2009, \apj, 694, 302, \dodoi{10.1088/0004-637X/694/1/302}

\bibitem[{{Driver} {et~al.}(2022){Driver}, {Bellstedt}, {Robotham}, {Baldry}, {Davies}, {Liske}, {Obreschkow}, {Taylor}, {Wright}, {Alpaslan}, {Bamford}, {Bauer}, {Bland-Hawthorn}, {Bilicki}, {Bravo}, {Brough}, {Casura}, {Cluver}, {Colless}, {Conselice}, {Croom}, {de Jong}, {D'Eugenio}, {De Propris}, {Dogruel}, {Drinkwater}, {Dvornik}, {Farrow}, {Frenk}, {Giblin}, {Graham}, {Grootes}, {Gunawardhana}, {Hashemizadeh}, {H{\"a}u{\ss}ler}, {Heymans}, {Hildebrandt}, {Holwerda}, {Hopkins}, {Jarrett}, {Heath Jones}, {Kelvin}, {Koushan}, {Kuijken}, {Lara-L{\'o}pez}, {Lange}, {L{\'o}pez-S{\'a}nchez}, {Loveday}, {Mahajan}, {Meyer}, {Moffett}, {Napolitano}, {Norberg}, {Owers}, {Radovich}, {Raouf}, {Peacock}, {Phillipps}, {Pimbblet}, {Popescu}, {Said}, {Sansom}, {Seibert}, {Sutherland}, {Thorne}, {Tuffs}, {Turner}, {van der Wel}, {van Kampen}, \& {Wilkins}}]{Driver2022}
{Driver}, S.~P., {Bellstedt}, S., {Robotham}, A. S.~G., {et~al.} 2022, \mnras, 513, 439, \dodoi{10.1093/mnras/stac472}

\bibitem[{{Dunn} {et~al.}(2018){Dunn}, {Bellovary}, {Holley-Bockelmann}, {Christensen}, \& {Quinn}}]{Dunn2018}
{Dunn}, G., {Bellovary}, J., {Holley-Bockelmann}, K., {Christensen}, C., \& {Quinn}, T. 2018, \apj, 861, 39, \dodoi{10.3847/1538-4357/aac7c2}

\bibitem[{{Elvis} {et~al.}(2002){Elvis}, {Risaliti}, \& {Zamorani}}]{Elvis2002}
{Elvis}, M., {Risaliti}, G., \& {Zamorani}, G. 2002, \apjl, 565, L75, \dodoi{10.1086/339197}

\bibitem[{{Fahrion} {et~al.}(2021){Fahrion}, {Lyubenova}, {van de Ven}, {Hilker}, {Leaman}, {Falc{\'o}n-Barroso}, {Bittner}, {Coccato}, {Corsini}, {Gadotti}, {Iodice}, {McDermid}, {Mart{\'\i}n-Navarro}, {Pinna}, {Poci}, {Sarzi}, {de Zeeuw}, \& {Zhu}}]{Fahrion2021}
{Fahrion}, K., {Lyubenova}, M., {van de Ven}, G., {et~al.} 2021, arXiv e-prints, arXiv:2104.06412.
\newblock \doarXiv{2104.06412}

\bibitem[{{Frank} \& {Rees}(1976)}]{Frank1976}
{Frank}, J., \& {Rees}, M.~J. 1976, \mnras, 176, 633, \dodoi{10.1093/mnras/176.3.633}

\bibitem[{{French} {et~al.}(2020){French}, {Arcavi}, {Zabludoff}, {Stone}, {Hiramatsu}, {van Velzen}, {McCully}, \& {Jiang}}]{French2020a}
{French}, K.~D., {Arcavi}, I., {Zabludoff}, A.~I., {et~al.} 2020, \apj, 891, 93, \dodoi{10.3847/1538-4357/ab7450}

\bibitem[{{Fryer} {et~al.}(2001){Fryer}, {Woosley}, \& {Heger}}]{Fryer2001}
{Fryer}, C.~L., {Woosley}, S.~E., \& {Heger}, A. 2001, \apj, 550, 372, \dodoi{10.1086/319719}

\bibitem[{{Georgiev} \& {B{\"o}ker}(2014)}]{Georgiev2014}
{Georgiev}, I.~Y., \& {B{\"o}ker}, T. 2014, \mnras, 441, 3570, \dodoi{10.1093/mnras/stu797}

\bibitem[{{Gezari}(2021)}]{Gezari2021}
{Gezari}, S. 2021, \araa, 59, 21, \dodoi{10.1146/annurev-astro-111720-030029}

\bibitem[{{Gonzaga} {et~al.}(2010){Gonzaga}, {Biretta}, \& et~al.}]{Gonzaga2010}
{Gonzaga}, S., {Biretta}, J., \& et~al. 2010, HST WFPC2 Data Handbook, v. 5.0, ed. edn., STScI, Baltimore

\bibitem[{{Greene} {et~al.}(2020){Greene}, {Strader}, \& {Ho}}]{Greene2020}
{Greene}, J.~E., {Strader}, J., \& {Ho}, L.~C. 2020, \araa, 58, 257, \dodoi{10.1146/annurev-astro-032620-021835}

\bibitem[{{Hills}(1975)}]{Hills1975}
{Hills}, J.~G. 1975, \nat, 254, 295, \dodoi{10.1038/254295a0}

\bibitem[{{Hirano} {et~al.}(2014){Hirano}, {Hosokawa}, {Yoshida}, {Umeda}, {Omukai}, {Chiaki}, \& {Yorke}}]{Hirano2014}
{Hirano}, S., {Hosokawa}, T., {Yoshida}, N., {et~al.} 2014, \apj, 781, 60, \dodoi{10.1088/0004-637X/781/2/60}

\bibitem[{{Ho}(2009)}]{Ho2009}
{Ho}, L.~C. 2009, \apj, 699, 626, \dodoi{10.1088/0004-637X/699/1/626}

\bibitem[{{Hopman} \& {Alexander}(2005)}]{Hopman2005}
{Hopman}, C., \& {Alexander}, T. 2005, \apj, 629, 362, \dodoi{10.1086/431475}

\bibitem[{{Hoyer} {et~al.}(2021){Hoyer}, {Neumayer}, {Georgiev}, {Seth}, \& {Greene}}]{Hoyer2021}
{Hoyer}, N., {Neumayer}, N., {Georgiev}, I.~Y., {Seth}, A.~C., \& {Greene}, J.~E. 2021, \mnras, 507, 3246, \dodoi{10.1093/mnras/stab2277}

\bibitem[{{Hoyer} {et~al.}(2023){Hoyer}, {Neumayer}, {Seth}, {Georgiev}, \& {Greene}}]{Hoyer2023a}
{Hoyer}, N., {Neumayer}, N., {Seth}, A.~C., {Georgiev}, I.~Y., \& {Greene}, J.~E. 2023, \mnras, 520, 4664, \dodoi{10.1093/mnras/stad220}

\bibitem[{{Inayoshi} {et~al.}(2020){Inayoshi}, {Visbal}, \& {Haiman}}]{Inayoshi2020}
{Inayoshi}, K., {Visbal}, E., \& {Haiman}, Z. 2020, \araa, 58, 27, \dodoi{10.1146/annurev-astro-120419-014455}

\bibitem[{{Johnson} {et~al.}(2011){Johnson}, {Khochfar}, {Greif}, \& {Durier}}]{Johnson2011}
{Johnson}, J.~L., {Khochfar}, S., {Greif}, T.~H., \& {Durier}, F. 2011, \mnras, 410, 919, \dodoi{10.1111/j.1365-2966.2010.17491.x}

\bibitem[{{Kormendy} \& {Ho}(2013)}]{Kormendy2013}
{Kormendy}, J., \& {Ho}, L.~C. 2013, \araa, 51, 511, \dodoi{10.1146/annurev-astro-082708-101811}

\bibitem[{{Kroupa}(2001)}]{Kroupa2001}
{Kroupa}, P. 2001, \mnras, 322, 231, \dodoi{10.1046/j.1365-8711.2001.04022.x}

\bibitem[{{Lauer} {et~al.}(1995){Lauer}, {Ajhar}, {Byun}, {Dressler}, {Faber}, {Grillmair}, {Kormendy}, {Richstone}, \& {Tremaine}}]{Lauer1995}
{Lauer}, T.~R., {Ajhar}, E.~A., {Byun}, Y.~I., {et~al.} 1995, \aj, 110, 2622, \dodoi{10.1086/117719}

\bibitem[{{Lauer} {et~al.}(2005){Lauer}, {Faber}, {Gebhardt}, {Richstone}, {Tremaine}, {Ajhar}, {Aller}, {Bender}, {Dressler}, {Filippenko}, {Green}, {Grillmair}, {Ho}, {Kormendy}, {Magorrian}, {Pinkney}, \& {Siopis}}]{Lauer2005}
{Lauer}, T.~R., {Faber}, S.~M., {Gebhardt}, K., {et~al.} 2005, \aj, 129, 2138, \dodoi{10.1086/429565}

\bibitem[{{Lauer} {et~al.}(2007){Lauer}, {Gebhardt}, {Faber}, {Richstone}, {Tremaine}, {Kormendy}, {Aller}, {Bender}, {Dressler}, {Filippenko}, {Green}, \& {Ho}}]{Lauer2007}
{Lauer}, T.~R., {Gebhardt}, K., {Faber}, S.~M., {et~al.} 2007, \apj, 664, 226, \dodoi{10.1086/519229}

\bibitem[{{Lezhnin} \& {Vasiliev}(2015)}]{Lezhnin2015}
{Lezhnin}, K., \& {Vasiliev}, E. 2015, \apjl, 808, L5, \dodoi{10.1088/2041-8205/808/1/L5}

\bibitem[{{Lodato} \& {Natarajan}(2006)}]{Lodato2006}
{Lodato}, G., \& {Natarajan}, P. 2006, \mnras, 371, 1813, \dodoi{10.1111/j.1365-2966.2006.10801.x}

\bibitem[{{Loeb} \& {Rasio}(1994)}]{Loeb1994}
{Loeb}, A., \& {Rasio}, F.~A. 1994, \apj, 432, 52, \dodoi{10.1086/174548}

\bibitem[{{Magorrian} \& {Tremaine}(1999)}]{Magorrian1999}
{Magorrian}, J., \& {Tremaine}, S. 1999, \mnras, 309, 447, \dodoi{10.1046/j.1365-8711.1999.02853.x}

\bibitem[{{McDermid} {et~al.}(2015){McDermid}, {Alatalo}, {Blitz}, {Bournaud}, {Bureau}, {Cappellari}, {Crocker}, {Davies}, {Davis}, {de Zeeuw}, {Duc}, {Emsellem}, {Khochfar}, {Krajnovi{\'c}}, {Kuntschner}, {Morganti}, {Naab}, {Oosterloo}, {Sarzi}, {Scott}, {Serra}, {Weijmans}, \& {Young}}]{McDermid2015}
{McDermid}, R.~M., {Alatalo}, K., {Blitz}, L., {et~al.} 2015, \mnras, 448, 3484, \dodoi{10.1093/mnras/stv105}

\bibitem[{{Miller} {et~al.}(2012){Miller}, {Gallo}, {Treu}, \& {Woo}}]{Miller2012}
{Miller}, B., {Gallo}, E., {Treu}, T., \& {Woo}, J.-H. 2012, \apj, 747, 57, \dodoi{10.1088/0004-637X/747/1/57}

\bibitem[{{Miniutti} {et~al.}(2019){Miniutti}, {Saxton}, {Giustini}, {Alexander}, {Fender}, {Heywood}, {Monageng}, {Coriat}, {Tzioumis}, {Read}, {Knigge}, {Gandhi}, {Pretorius}, \& {Ag{\'\i}s-Gonz{\'a}lez}}]{Miniutti2019}
{Miniutti}, G., {Saxton}, R.~D., {Giustini}, M., {et~al.} 2019, \nat, 573, 381, \dodoi{10.1038/s41586-019-1556-x}

\bibitem[{{Neumayer} {et~al.}(2020){Neumayer}, {Seth}, \& {B{\"o}ker}}]{Neumayer2020}
{Neumayer}, N., {Seth}, A., \& {B{\"o}ker}, T. 2020, \aapr, 28, 4, \dodoi{10.1007/s00159-020-00125-0}

\bibitem[{{Nguyen} {et~al.}(2018){Nguyen}, {Seth}, {Neumayer}, {Kamann}, {Voggel}, {Cappellari}, {Picotti}, {Nguyen}, {B{\"o}ker}, {Debattista}, {Caldwell}, {McDermid}, {Bastian}, {Ahn}, \& {Pechetti}}]{Nguyen2018}
{Nguyen}, D.~D., {Seth}, A.~C., {Neumayer}, N., {et~al.} 2018, \apj, 858, 118, \dodoi{10.3847/1538-4357/aabe28}

\bibitem[{{Ohlson} {et~al.}(2023){Ohlson}, {Seth}, {Gallo}, {Baldassare}, \& {Greene}}]{Ohlson2023}
{Ohlson}, D., {Seth}, A.~C., {Gallo}, E., {Baldassare}, V.~F., \& {Greene}, J.~E. 2023, arXiv e-prints, arXiv:2309.05701, \dodoi{10.48550/arXiv.2309.05701}

\bibitem[{{O'Sullivan} {et~al.}(2001){O'Sullivan}, {Forbes}, \& {Ponman}}]{Osullivan2001}
{O'Sullivan}, E., {Forbes}, D.~A., \& {Ponman}, T.~J. 2001, \mnras, 328, 461, \dodoi{10.1046/j.1365-8711.2001.04890.x}

\bibitem[{{Pechetti} {et~al.}(2017){Pechetti}, {Seth}, {Cappellari}, {McDermid}, {den Brok}, {Mieske}, \& {Strader}}]{Pechetti2017}
{Pechetti}, R., {Seth}, A., {Cappellari}, M., {et~al.} 2017, \apj, 850, 15, \dodoi{10.3847/1538-4357/aa9021}

\bibitem[{{Pechetti} {et~al.}(2020){Pechetti}, {Seth}, {Neumayer}, {Georgiev}, {Kacharov}, \& {den Brok}}]{Pechetti2020}
{Pechetti}, R., {Seth}, A., {Neumayer}, N., {et~al.} 2020, \apj, 900, 32, \dodoi{10.3847/1538-4357/abaaa7}

\bibitem[{{Pfister} {et~al.}(2020){Pfister}, {Volonteri}, {Dai}, \& {Colpi}}]{Pfister2020}
{Pfister}, H., {Volonteri}, M., {Dai}, J.~L., \& {Colpi}, M. 2020, \mnras, 497, 2276, \dodoi{10.1093/mnras/staa1962}

\bibitem[{Polkas {et~al.}(2023)Polkas, Bonoli, Bortolas, Izquierdo-Villalba, Sesana, Broggi, Hoyer, \& Spinoso}]{Polkas2023}
Polkas, M., Bonoli, S., Bortolas, E., {et~al.} 2023, Demographics of Tidal Disruption Events with L-Galaxies: I. Volumetric TDE rates and the abundance of Nuclear Star Clusters.
\newblock \doarXiv{2312.13242}

\bibitem[{{Portegies Zwart} \& {McMillan}(2002)}]{Portegies-Zwart2002}
{Portegies Zwart}, S.~F., \& {McMillan}, S. L.~W. 2002, \apj, 576, 899, \dodoi{10.1086/341798}

\bibitem[{{Qunbar} \& {Stone}(2023)}]{Qunbar2023}
{Qunbar}, I., \& {Stone}, N.~C. 2023, arXiv e-prints, arXiv:2304.13062, \dodoi{10.48550/arXiv.2304.13062}

\bibitem[{{Rees}(1988)}]{Rees1988}
{Rees}, M.~J. 1988, \nat, 333, 523, \dodoi{10.1038/333523a0}

\bibitem[{{Reines}(2022)}]{Reines2022}
{Reines}, A.~E. 2022, Nature Astronomy, 6, 26, \dodoi{10.1038/s41550-021-01556-0}

\bibitem[{{Reines} \& {Volonteri}(2015)}]{Reines2015}
{Reines}, A.~E., \& {Volonteri}, M. 2015, \apj, 813, 82, \dodoi{10.1088/0004-637X/813/2/82}

\bibitem[{{Ricarte} \& {Natarajan}(2018)}]{Ricarte2018}
{Ricarte}, A., \& {Natarajan}, P. 2018, \mnras, 481, 3278, \dodoi{10.1093/mnras/sty2448}

\bibitem[{{Rizzuto} {et~al.}(2021){Rizzuto}, {Naab}, {Spurzem}, {Giersz}, {Ostriker}, {Stone}, {Wang}, {Berczik}, \& {Rampp}}]{Rizzuto2021}
{Rizzuto}, F.~P., {Naab}, T., {Spurzem}, R., {et~al.} 2021, \mnras, 501, 5257, \dodoi{10.1093/mnras/staa3634}

\bibitem[{{S{\'a}nchez-Janssen} {et~al.}(2019){S{\'a}nchez-Janssen}, {C{\^o}t{\'e}}, {Ferrarese}, {Peng}, {Roediger}, {Blakeslee}, {Emsellem}, {Puzia}, {Spengler}, {Taylor}, {{\'A}lamo-Mart{\'\i}nez}, {Boselli}, {Cantiello}, {Cuillandre}, {Duc}, {Durrell}, {Gwyn}, {MacArthur}, {Lan{\c{c}}on}, {Lim}, {Liu}, {Mei}, {Miller}, {Mu{\~n}oz}, {Mihos}, {Paudel}, {Powalka}, \& {Toloba}}]{Sanchez-Janssen2019}
{S{\'a}nchez-Janssen}, R., {C{\^o}t{\'e}}, P., {Ferrarese}, L., {et~al.} 2019, \apj, 878, 18, \dodoi{10.3847/1538-4357/aaf4fd}

\bibitem[{{Schlafly} \& {Finkbeiner}(2011)}]{Schlafly2011}
{Schlafly}, E.~F., \& {Finkbeiner}, D.~P. 2011, \apj, 737, 103, \dodoi{10.1088/0004-637X/737/2/103}

\bibitem[{{Schlegel} {et~al.}(1998){Schlegel}, {Finkbeiner}, \& {Davis}}]{Schlegel1998}
{Schlegel}, D.~J., {Finkbeiner}, D.~P., \& {Davis}, M. 1998, \apj, 500, 525, \dodoi{10.1086/305772}

\bibitem[{{She} {et~al.}(2017){She}, {Ho}, \& {Feng}}]{She2017}
{She}, R., {Ho}, L.~C., \& {Feng}, H. 2017, \apj, 835, 223, \dodoi{10.3847/1538-4357/835/2/223}

\bibitem[{{Shi} {et~al.}(2021){Shi}, {Grudi{\'c}}, \& {Hopkins}}]{Shi2021}
{Shi}, Y., {Grudi{\'c}}, M.~Y., \& {Hopkins}, P.~F. 2021, \mnras, 505, 2753, \dodoi{10.1093/mnras/stab1470}

\bibitem[{{Shvartzvald} {et~al.}(2023){Shvartzvald}, {Waxman}, {Gal-Yam}, {Ofek}, {Ben-Ami}, {Berge}, {Kowalski}, {B{\"u}hler}, {Worm}, {Rhoads}, {Arcavi}, {Maoz}, {Polishook}, {Stone}, {Trakhtenbrot}, {Ackermann}, {Aharonson}, {Birnholtz}, {Chelouche}, {Guetta}, {Hallakoun}, {Horesh}, {Kushnir}, {Mazeh}, {Nordin}, {Ofir}, {Ohm}, {Parsons}, {Pe'er}, {Perets}, {Perdelwitz}, {Poznanski}, {Sadeh}, {Sagiv}, {Shahaf}, {Soumagnac}, {Tal-Or}, {Van Santen}, {Zackay}, {Guttman}, {Rekhi}, {Townsend}, {Weinstein}, \& {Wold}}]{Ben-Ami2023}
{Shvartzvald}, Y., {Waxman}, E., {Gal-Yam}, A., {et~al.} 2023, arXiv e-prints, arXiv:2304.14482, \dodoi{10.48550/arXiv.2304.14482}

\bibitem[{{Soltan}(1982)}]{Soltan1982}
{Soltan}, A. 1982, \mnras, 200, 115, \dodoi{10.1093/mnras/200.1.115}

\bibitem[{{Spera} \& {Mapelli}(2017)}]{Spera2017}
{Spera}, M., \& {Mapelli}, M. 2017, \mnras, 470, 4739, \dodoi{10.1093/mnras/stx1576}

\bibitem[{{Stone} {et~al.}(2018){Stone}, {Generozov}, {Vasiliev}, \& {Metzger}}]{Stone2018}
{Stone}, N.~C., {Generozov}, A., {Vasiliev}, E., \& {Metzger}, B.~D. 2018, \mnras, 480, 5060, \dodoi{10.1093/mnras/sty2045}

\bibitem[{{Stone} {et~al.}(2017){Stone}, {K{\"u}pper}, \& {Ostriker}}]{Stone2017}
{Stone}, N.~C., {K{\"u}pper}, A. H.~W., \& {Ostriker}, J.~P. 2017, \mnras, 467, 4180, \dodoi{10.1093/mnras/stx097}

\bibitem[{{Stone} \& {Metzger}(2016)}]{StoneMetzger2016}
{Stone}, N.~C., \& {Metzger}, B.~D. 2016, \mnras, 455, 859, \dodoi{10.1093/mnras/stv2281}

\bibitem[{{Stone} \& {van Velzen}(2016)}]{Stone2016}
{Stone}, N.~C., \& {van Velzen}, S. 2016, \apjl, 825, L14, \dodoi{10.3847/2041-8205/825/1/L14}

\bibitem[{{Taylor} {et~al.}(2011){Taylor}, {Hopkins}, {Baldry}, {Brown}, {Driver}, {Kelvin}, {Hill}, {Robotham}, {Bland-Hawthorn}, {Jones}, {Sharp}, {Thomas}, {Liske}, {Loveday}, {Norberg}, {Peacock}, {Bamford}, {Brough}, {Colless}, {Cameron}, {Conselice}, {Croom}, {Frenk}, {Gunawardhana}, {Kuijken}, {Nichol}, {Parkinson}, {Phillipps}, {Pimbblet}, {Popescu}, {Prescott}, {Sutherland}, {Tuffs}, {van Kampen}, \& {Wijesinghe}}]{Taylor2011}
{Taylor}, E.~N., {Hopkins}, A.~M., {Baldry}, I.~K., {et~al.} 2011, \mnras, 418, 1587, \dodoi{10.1111/j.1365-2966.2011.19536.x}

\bibitem[{{van den Bosch}(2016)}]{van-den-Bosch2016}
{van den Bosch}, R. C.~E. 2016, \apj, 831, 134, \dodoi{10.3847/0004-637X/831/2/134}

\bibitem[{{van Velzen}(2018)}]{vanVelzen2018}
{van Velzen}, S. 2018, \apj, 852, 72, \dodoi{10.3847/1538-4357/aa998e}

\bibitem[{{V{\'e}ron-Cetty} \& {V{\'e}ron}(2010)}]{Veron-Cetty2010}
{V{\'e}ron-Cetty}, M.~P., \& {V{\'e}ron}, P. 2010, \aap, 518, A10, \dodoi{10.1051/0004-6361/201014188}

\bibitem[{{Volonteri} {et~al.}(2008){Volonteri}, {Lodato}, \& {Natarajan}}]{Volonteri2008}
{Volonteri}, M., {Lodato}, G., \& {Natarajan}, P. 2008, \mnras, 383, 1079, \dodoi{10.1111/j.1365-2966.2007.12589.x}

\bibitem[{{Wang} \& {Merritt}(2004)}]{Wang2004}
{Wang}, J., \& {Merritt}, D. 2004, \apj, 600, 149, \dodoi{10.1086/379767}

\bibitem[{{Wevers} {et~al.}(2022){Wevers}, {Pasham}, {Jalan}, {Rakshit}, \& {Arcodia}}]{Wevers2022}
{Wevers}, T., {Pasham}, D.~R., {Jalan}, P., {Rakshit}, S., \& {Arcodia}, R. 2022, \aap, 659, L2, \dodoi{10.1051/0004-6361/202243143}

\bibitem[{{Wise} {et~al.}(2019){Wise}, {Regan}, {O'Shea}, {Norman}, {Downes}, \& {Xu}}]{Wise2019}
{Wise}, J.~H., {Regan}, J.~A., {O'Shea}, B.~W., {et~al.} 2019, \nat, 566, 85, \dodoi{10.1038/s41586-019-0873-4}

\bibitem[{{Yao} {et~al.}(2023){Yao}, {Ravi}, {Gezari}, {van Velzen}, {Lu}, {Schulze}, {Somalwar}, {Kulkarni}, {Hammerstein}, {Nicholl}, {Graham}, {Perley}, {Cenko}, {Stein}, {Ricarte}, {Chadayammuri}, {Quataert}, {Bellm}, {Bloom}, {Dekany}, {Drake}, {Groom}, {Mahabal}, {Prince}, {Riddle}, {Rusholme}, {Sharma}, {Sollerman}, \& {Yan}}]{Yao2023}
{Yao}, Y., {Ravi}, V., {Gezari}, S., {et~al.} 2023, \apjl, 955, L6, \dodoi{10.3847/2041-8213/acf216}

\bibitem[{{Yoshida} {et~al.}(2006){Yoshida}, {Omukai}, {Hernquist}, \& {Abel}}]{Yoshida2006}
{Yoshida}, N., {Omukai}, K., {Hernquist}, L., \& {Abel}, T. 2006, \apj, 652, 6, \dodoi{10.1086/507978}

\bibitem[{{Zocchi} {et~al.}(2019){Zocchi}, {Gieles}, \& {H{\'e}nault-Brunet}}]{Zocchi2019}
{Zocchi}, A., {Gieles}, M., \& {H{\'e}nault-Brunet}, V. 2019, \mnras, 482, 4713, \dodoi{10.1093/mnras/sty1508}

\end{thebibliography}
\bibliographystyle{aasjournal}

\begin{longrotatetable}

\begin{deluxetable*}{cccccccccccccccccc}
\tablecolumns{18}
\tabletypesize{\scriptsize}
\tablecaption{ \label{tab:table_1}}
\tablehead{
    \colhead{Name} & \colhead{RA} & \colhead{DEC} & \colhead{Type} & \colhead{log($M_{\rm gal}$)} & \colhead{Dist.} & \colhead{$(g-i)_{\rm gal}$} & \colhead{$(g-i)_{\rm nuc}$} & \colhead{src$_{g-i}$} & \colhead{NSC} & \colhead{$M/L_i$} & \colhead{src$_{M/L}$} & \colhead{$\gamma$} & \colhead{log($\rho_{\rm 5pc}$)} & \colhead{log($M_{\rm BH}$)} & \colhead{src$_{M_{\rm BH}}$} & \colhead{$q$} & \colhead{Sample}  \\
    \colhead{(1)} & \colhead{(2)} & \colhead{(3)} & \colhead{(4)} & \colhead{(5)} & \colhead{(6)} & \colhead{(7)} & \colhead{(8)} & \colhead{(9)} & \colhead{(10)} & \colhead{(11)} & \colhead{(12)} & \colhead{(13)} & \colhead{(14)} & \colhead{(15)} & \colhead{(16)} & \colhead{(17)} & \colhead{(18)}
}

\startdata
\hline
BTS 076 & 179.68375 & 27.585 & Late & 7.51 & 12.59 & 0.82 & 0.34 & 1 & Y & 0.36 & C-M/L & -2.41 & 0.41 & 4.03 & 5 & 0.85 & H23 \\ 
BTS 109 & 184.29208 & 47.06361 & Late & 6.58 & 13.8 & 0.0 & 0.54 & 1 & Y & 0.5 & C-M/L & -2.97 & 0.15 & 3.12 & 5 & 0.833 & H23 \\ 
DDO 084 & 160.67458 & 34.44889 & Late & 8.53 & 9.95 & 0.7 & 0.75 & 1 & Y & 0.7 & C-M/L & -2.9 & 1.1 & 5.03 & 5 & 0.872 & H23 \\ 
ESO 274-1 & 228.55767 & -46.80794 & Late & 9.12 & 2.79 & 1.18 & 1.08 & 2 & Y & 1.2 & C-M/L & -2.71 & 2.3 & 5.61 & 5 & 0.993 & P20 \\ 
IC 5052 & 313.02321 & -69.20164 & Late & 9.18 & 5.5 & 0.97 & 1.41 & 2 & Y & 2.01 & C-M/L & -2.42 & 2.4 & 5.67 & 5 & 0.967 & P20 \\ 
LeG 09 & 160.64417 & 12.15056 & Early & 6.99 & 10.19 & 0.92 & 1.13 & 1 & Y & 1.29 & C-M/L & -2.15 & 1.24 & 3.26 & 5 & 0.929 & H23 \\ 
NGC 0584 & 22.83633 & -6.86836 & Early & 10.56 & 20.0 & 1.1 & 1.32 & 3 & N & 1.75 & C-M/L & -1.38 & 3.74 & 8.11 & 3 & 0.654 & L05 \\ 
NGC 0596 & 23.21627 & -7.03154 & Early & 10.46 & 21.83 & 1.09 & 1.2 & 3 & Y & 1.44 & C-M/L & -1.45 & 3.75 & 7.89 & 4 & 0.877 & L05 \\ 
NGC 0821 & 32.08767 & 10.99475 & Early & 10.7 & 23.2 & 1.17 & 1.32 & 3 & N & 1.61 & SED & -1.36 & 4.03 & 8.22 & 1 & 0.63 & L05 \\ 
NGC 1374 & 53.81917 & -35.22611 & Early & 10.3 & 19.7 & 1.09 & 1.26 & 3 & N & 1.59 & C-M/L & -0.4 & 3.64 & 8.77 & 3 & 0.964 & L05 \\ 
NGC 1399 & 54.62217 & -35.45019 & Early & 11.16 & 21.1 & 1.24 & 1.35 & 3 & Y & 1.83 & C-M/L & -1.12 & 2.63 & 8.94 & 3 & 0.922 & L05 \\ 
NGC 1427 & 55.58083 & -35.39278 & Early & 10.39 & 19.4 & 1.12 & 1.24 & 3 & Y & 1.53 & C-M/L & -2.24 & 3.7 & 7.97 & 4 & 0.713 & L05 \\ 
NGC 1439 & 56.21 & -21.92222 & Early & 10.49 & 26.7 & 1.12 & 1.34 & 3 & Y & 1.8 & C-M/L & -2.23 & 4.02 & 7.86 & 4 & 0.726 & L05 \\ 
NGC 2300 & 113.09092 & 85.70894 & Early & 10.81 & 27.55 & 1.33 & 1.32 & 3 & N & 1.76 & C-M/L & -0.82 & 2.47 & 9.09 & 4 & 0.804 & L05 \\ 
NGC 2434 & 113.71318 & -69.28415 & Early & 10.58 & 21.8 & 1.1 & 1.34 & 3 & Y & 1.82 & C-M/L & -1.15 & 3.98 & 8.29 & 4 & 0.917 & L05 \\ 
NGC 2778 & 138.10158 & 35.0275 & Early & 9.94 & 23.2 & 1.16 & 1.3 & 3 & Y & 2.07 & SED & -2.05 & 4.29 & 7.15 & 3 & 0.862 & P17 \\ 
NGC 2787 & 139.82749 & 69.20325 & Late & 9.95 & 7.48 & 1.28 & 1.41 & 2 & Y & 2.02 & C-M/L & -1.31 & 3.7 & 7.61 & 3 & 1.0 & P20 \\ 
NGC 2903 & 143.04212 & 21.50083 & Late & 10.4 & 8.87 & 1.04 & 1.52 & 2 & Y & 2.42 & C-M/L & -1.65 & 3.45 & 7.06 & 1 & 0.991 & P20 \\ 
NGC 3115B & 151.42333 & -7.9815 & Early & 9.03 & 9.7 & 1.06 & 1.02 & 2 & Y & 1.08 & C-M/L & -1.85 & 2.9 & 4.69 & 4 & 0.981 & P20 \\ 
NGC 3274 & 158.07196 & 27.66878 & Late & 8.22 & 7.98 & 0.4 & 0.82 & 2 & Y & 0.78 & C-M/L & -2.84 & 1.92 & 4.73 & 5 & 0.971 & P20 \\ 
NGC 3344 & 160.87979 & 24.92222 & Late & 9.9 & 9.82 & 0.89 & 0.52 & 2 & Y & 0.48 & C-M/L & -1.62 & 3.49 & 6.54 & 4 & 0.987 & P20 \\ 
NGC 3379 & 161.95729 & 12.58183 & Early & 10.63 & 11.32 & 1.27 & 1.32 & 3 & N & 1.93 & SED & -1.08 & 3.51 & 8.62 & 3 & 0.898 & L05 \\ 
NGC 3384 & 162.07042 & 12.62858 & Early & 10.13 & 9.42 & 1.19 & 0.17 & 4 & Y & 2.43 & SED & -1.98 & 3.75 & 7.04 & 3 & 0.585 & L05 \\ 
NGC 3412 & 162.7221 & 13.41216 & Early & 10.01 & 11.32 & 1.09 & 1.15 & 3 & Y & 1.18 & SED & -2.23 & 2.87 & 7.31 & 4 & 0.318 & P17 \\ 
NGC 3522 & 166.66859 & 20.0856 & Early & 9.69 & 25.2 & 1.03 & 1.2 & 3 & Y & 1.33 & SED & -1.94 & 3.73 & 7.1 & 4 & 0.631 & P17 \\ 
NGC 3585 & 168.32129 & -26.75499 & Early & 10.93 & 20.4 & 1.22 & 1.3 & 3 & N & 1.7 & C-M/L & -1.23 & 3.8 & 8.52 & 3 & 0.717 & L05 \\ 
NGC 3607 & 169.22833 & 18.05111 & Early & 10.87 & 22.8 & 1.2 & 1.43 & 3 & N & 2.02 & SED & -0.95 & 3.48 & 8.15 & 3 & 0.871 & L05 \\ 
NGC 3608 & 169.24612 & 18.1485 & Early & 10.45 & 22.9 & 1.16 & 1.28 & 3 & N & 1.84 & SED & -0.98 & 3.74 & 8.66 & 3 & 0.892 & L05 \\ 
NGC 3610 & 169.60587 & 58.78631 & Early & 10.8 & 34.8 & 1.05 & 0.88 & 4 & N & 1.24 & SED & -1.77 & 4.05 & 8.1 & 4 & 0.563 & L05 \\ 
NGC 3640 & 170.27858 & 3.23506 & Early & 10.9 & 27.0 & 1.11 & 1.25 & 3 & N & 1.77 & SED & -0.23 & 3.56 & 7.89 & 3 & 0.89 & L05 \\ 
NGC 3945 & 178.30746 & 60.67544 & Early & 10.54 & 21.51 & 1.26 & 1.33 & 3 & Y & 1.97 & SED & -2.4 & 4.11 & 6.94 & 3 & 0.809 & L05 \\ 
NGC 4026 & 179.85492 & 50.96139 & Early & 10.06 & 13.6 & 1.11 & 1.28 & 3 & Y & 1.92 & SED & -1.67 & 3.93 & 8.26 & 3 & 0.531 & L05 \\ 
NGC 4242 & 184.37574 & 45.6193 & Late & 9.1 & 7.9 & 0.68 & 0.75 & 2 & Y & 0.7 & C-M/L & -3.07 & 1.67 & 5.59 & 5 & 0.997 & P20 \\ 
NGC 4262 & 184.87751 & 14.87756 & Early & 9.97 & 15.42 & 1.18 & 0.98 & 4 & Y & 2.16 & SED & -1.32 & 4.17 & 8.61 & 4 & 0.904 & P17 \\ 
NGC 4291 & 185.07333 & 75.37083 & Early & 10.5 & 26.3 & 1.24 & 1.26 & 3 & N & 1.59 & C-M/L & -0.56 & 2.91 & 8.99 & 3 & 0.74 & L05 \\ 
NGC 4342 & 185.91252 & 7.05395 & Early & 9.93 & 19.5 & 1.21 & 1.29 & 3 & Y & 2.07 & SED & -1.99 & 4.48 & 8.65 & 3 & 0.808 & P17 \\ 
NGC 4365 & 186.11808 & 7.31783 & Early & 11.09 & 23.34 & 1.2 & 1.3 & 3 & Y & 2.01 & SED & -0.5 & 2.97 & 8.89 & 4 & 0.714 & L05 \\ 
NGC 4377 & 186.30135 & 14.76222 & Early & 10.02 & 17.78 & 1.14 & 1.22 & 3 & Y & 1.82 & SED & -2.29 & 3.16 & 7.62 & 4 & 0.328 & P17 \\ 
NGC 4379 & 186.31139 & 15.60742 & Early & 9.92 & 15.85 & 1.15 & 1.24 & 3 & Y & 1.82 & SED & -2.32 & 3.96 & 7.48 & 4 & 0.8 & P17 \\ 
NGC 4382 & 186.35096 & 18.19081 & Early & 11.02 & 17.86 & 1.08 & 1.06 & 3 & N & 1.12 & SED & -0.57 & 2.99 & 7.11 & 3 & 0.738 & L05 \\ 
NGC 4387 & 186.4237 & 12.81043 & Early & 9.84 & 17.95 & 1.16 & 1.19 & 3 & Y & 1.83 & SED & -2.15 & 3.75 & 7.16 & 4 & 1.0 & P17 \\ 
NGC 4434 & 186.90284 & 8.15432 & Early & 10.03 & 22.39 & 1.13 & 1.23 & 3 & Y & 1.96 & SED & -1.87 & 4.17 & 7.85 & 3 & 1.0 & P17 \\ 
NGC 4458 & 187.23984 & 13.2419 & Early & 9.73 & 16.37 & 1.1 & 1.16 & 3 & N & 1.83 & SED & -0.62 & 3.79 & 7.12 & 4 & 0.694 & P17 \\ 
NGC 4464 & 187.33908 & 8.15642 & Early & 9.58 & 15.85 & 1.16 & 1.16 & 4 & Y & 1.35 & C-M/L & -1.53 & 3.7 & 7.57 & 4 & 0.903 & L95 \\ 
NGC 4467 & 187.37646 & 7.99397 & Early & 9.04 & 16.5 & 1.24 & 1.51 & 3 & Y & 2.39 & C-M/L & -1.33 & 3.58 & 6.41 & 4 & 0.977 & L95 \\ 
NGC 4472 & 187.44483 & 7.99997 & Early & 11.3 & 17.14 & 1.23 & 1.34 & 3 & Y & 2.04 & SED & -1.46 & 2.76 & 9.4 & 3 & 0.912 & L05 \\ 
NGC 4473 & 187.45396 & 13.42947 & Early & 10.49 & 15.28 & 1.18 & 1.29 & 3 & N & 2.13 & SED & -0.81 & 3.52 & 7.95 & 3 & 0.531 & L05 \\ 
NGC 4474 & 187.47316 & 14.0686 & Early & 9.97 & 15.56 & 1.17 & 1.17 & 3 & Y & 1.53 & SED & -2.18 & 3.61 & 6.97 & 4 & 0.695 & P17 \\ 
NGC 4478 & 187.57304 & 12.32786 & Early & 10.08 & 16.98 & 1.11 & 1.21 & 3 & Y & 1.97 & SED & -1.79 & 3.61 & 7.81 & 4 & 0.689 & L05 \\ 
NGC 4483 & 187.66936 & 9.01567 & Early & 9.72 & 16.75 & 1.12 & 1.21 & 3 & Y & 1.7 & SED & -1.66 & 3.87 & 7.16 & 4 & 0.875 & P17 \\ 
NGC 4486B & 187.63258 & 12.49053 & Early & 9.33 & 16.29 & 1.21 & 1.28 & 3 & N & 1.63 & C-M/L & -1.0 & 3.46 & 8.78 & 3 & 0.55 & L05 \\ 
NGC 4489 & 187.71769 & 16.75882 & Early & 9.65 & 15.42 & 1.07 & 1.09 & 3 & Y & 0.9 & SED & -1.35 & 3.42 & 6.13 & 4 & 1.0 & P17 \\ 
NGC 4494 & 187.85017 & 25.77469 & Early & 10.68 & 16.9 & 1.14 & 1.36 & 3 & N & 1.73 & SED & -1.44 & 4.24 & 7.91 & 4 & 0.775 & L05 \\ 
NGC 4517 & 188.18996 & 0.11503 & Late & 9.86 & 8.36 & 1.14 & 0.57 & 2 & Y & 0.52 & C-M/L & -2.72 & 2.12 & 5.97 & 4 & 0.889 & P20 \\ 
NGC 4528 & 188.52535 & 11.32123 & Early & 9.85 & 15.85 & 1.11 & 1.16 & 3 & Y & 1.45 & SED & -1.55 & 3.71 & 7.24 & 4 & 0.76 & P17 \\ 
NGC 4551 & 188.90816 & 12.26398 & Early & 9.86 & 16.14 & 1.16 & 1.22 & 3 & Y & 1.91 & SED & -1.88 & 3.62 & 7.2 & 4 & 0.696 & P17 \\ 
NGC 4552 & 188.91667 & 12.55636 & Early & 10.7 & 15.85 & 1.2 & 1.35 & 3 & Y & 1.96 & SED & -2.15 & 3.52 & 8.7 & 3 & 0.862 & L05 \\ 
NGC 4589 & 189.355 & 74.19206 & Early & 10.72 & 22.4 & 1.3 & 1.43 & 3 & N & 2.1 & C-M/L & -0.79 & 3.33 & 8.67 & 4 & 0.601 & L05 \\ 
NGC 4592 & 189.82807 & -0.53201 & Late & 9.45 & 16.5 & 0.59 & 0.95 & 2 & Y & 0.97 & C-M/L & -2.99 & 2.33 & 5.93 & 5 & 0.962 & P20 \\ 
NGC 4600 & 190.09566 & 3.11775 & Early & 9.01 & 9.29 & 1.02 & 0.95 & 2 & Y & 0.97 & C-M/L & -2.02 & 2.83 & 6.59 & 4 & 0.994 & P20 \\ 
NGC 4605 & 189.99742 & 61.60919 & Late & 9.17 & 5.55 & 0.69 & 1.46 & 2 & Y & 2.2 & C-M/L & -2.9 & 2.97 & 5.4 & 4 & 0.996 & P20 \\ 
NGC 4612 & 190.38646 & 7.31488 & Early & 9.97 & 16.6 & 1.07 & 0.88 & 4 & Y & 0.81 & SED & -2.53 & 3.7 & 6.77 & 4 & 1.0 & P17 \\ 
NGC 4623 & 190.54459 & 7.67698 & Early & 9.71 & 17.38 & 1.1 & 1.2 & 3 & Y & 1.41 & SED & -2.85 & 3.58 & 6.68 & 4 & 0.903 & P17 \\ 
NGC 4638 & 190.69758 & 11.44249 & Early & 10.14 & 17.46 & 1.13 & 1.05 & 4 & Y & 1.84 & SED & -1.36 & 4.37 & 7.51 & 4 & 1.0 & P17 \\ 
NGC 4649 & 190.91746 & 11.55247 & Early & 11.18 & 17.3 & 1.26 & 1.34 & 3 & N & 2.01 & SED & -1.47 & 3.02 & 9.67 & 3 & 0.928 & L05 \\ 
NGC 4660 & 191.13325 & 11.19051 & Early & 9.93 & 15.0 & 1.05 & 1.28 & 3 & N & 2.06 & SED & -2.65 & 4.57 & 8.59 & 4 & 0.838 & P17 \\ 
NGC 4733 & 192.77823 & 10.91209 & Early & 9.83 & 17.4 & 1.04 & 1.03 & 3 & Y & 0.91 & SED & -3.98 & 3.24 & 6.12 & 4 & 1.0 & P17 \\ 
NGC 5011C & 198.29958 & -43.26556 & Early & 7.52 & 3.73 & 0.99 & 1.64 & 1 & Y & 2.95 & C-M/L & -2.63 & 0.98 & 3.96 & 5 & 0.772 & H23 \\ 
NGC 5055 & 198.95554 & 42.02928 & Late & 10.5 & 9.04 & 0.97 & 0.4 & 2 & Y & 0.4 & C-M/L & -1.66 & 3.58 & 8.92 & 1 & 0.992 & P20 \\ 
NGC 5068 & 199.72837 & -21.03911 & Late & 9.39 & 5.15 & 0.71 & 0.86 & 2 & Y & 0.83 & C-M/L & -1.84 & 2.27 & 5.88 & 5 & 0.973 & P20 \\ 
NGC 5195 & 202.49829 & 47.26614 & Early & 10.1 & 7.66 & 1.17 & 1.47 & 2 & Y & 2.23 & C-M/L & -1.77 & 4.09 & 7.57 & 4 & 0.966 & P20 \\ 
NGC 5236 & 204.25396 & -29.86542 & Late & 10.44 & 4.9 & 0.74 & 1.41 & 2 & Y & 2.04 & C-M/L & -2.54 & 3.95 & 6.91 & 5 & 0.972 & P20 \\ 
NGC 5238 & 203.67714 & 51.61368 & Late & 7.75 & 4.51 & 0.57 & 0.15 & 2 & Y & 0.27 & C-M/L & -2.94 & 1.59 & 4.27 & 5 & 0.996 & P20 \\ 
NGC 5457 & 210.80227 & 54.34895 & Late & 10.21 & 6.95 & 0.93 & 1.06 & 2 & Y & 1.15 & C-M/L & -1.81 & 3.05 & 6.41 & 1 & 0.975 & P20 \\ 
NGC 5557 & 214.6075 & 36.49331 & Early & 10.88 & 49.67 & 1.12 & 1.1 & 4 & N & 2.06 & SED & -0.72 & 3.39 & 9.16 & 4 & 0.911 & L05 \\ 
NGC 5576 & 215.265 & 3.27067 & Early & 10.58 & 25.5 & 1.09 & 1.19 & 3 & N & 1.78 & SED & -1.11 & 4.0 & 8.43 & 3 & 0.758 & L05 \\ 
NGC 5813 & 225.29717 & 1.70178 & Early & 11.08 & 32.17 & 1.2 & 1.36 & 3 & N & 1.87 & SED & -0.01 & 2.37 & 8.85 & 3 & 0.876 & L05 \\ 
NGC 5982 & 234.6665 & 59.35578 & Early & 10.97 & 40.4 & 1.17 & 1.24 & 3 & N & 1.54 & C-M/L & -0.53 & 2.68 & 8.89 & 4 & 0.969 & L05 \\ 
NGC 6503 & 267.36013 & 70.14437 & Late & 9.64 & 6.25 & 1.03 & 1.65 & 2 & Y & 2.97 & C-M/L & -2.32 & 3.06 & 6.3 & 1 & 1.0 & P20 \\ 
NGC 7457 & 345.24971 & 30.14489 & Early & 9.88 & 12.1 & 1.05 & 1.05 & 3 & Y & 0.71 & SED & -2.58 & 3.79 & 6.95 & 3 & 0.957 & P17 \\ 
NGC 7713 & 354.06246 & -37.93808 & Late & 8.97 & 7.8 & 0.39 & 0.2 & 2 & Y & 0.29 & C-M/L & -3.36 & 1.5 & 5.47 & 5 & 1.0 & P20 \\ 
NGC 7727 & 354.97446 & -12.29301 & Late & 10.67 & 23.37 & 1.09 & 1.41 & 3 & N & 2.04 & C-M/L & -1.29 & 3.68 & 7.59 & 4 & 0.738 & L05 \\ 
PGC 4310323 & 181.37917 & 31.07611 & Late & 6.61 & 6.43 & 0.74 & 0.95 & 1 & Y & 0.97 & C-M/L & -2.76 & 0.87 & 3.15 & 5 & 0.938 & H23 \\ 
UGC 07242 & 183.53083 & 66.09222 & Late & 7.75 & 5.45 & 0.0 & 0.97 & 1 & Y & 0.99 & C-M/L & -2.88 & 1.05 & 4.27 & 5 & 0.871 & H23 \\ 
VCC 1199 & 187.39571 & 8.05872 & Early & 8.6 & 16.5 & 1.3 & 1.23 & 3 & Y & 1.51 & C-M/L & -1.59 & 2.99 & 5.39 & 5 & 0.817 & L95 \\ 
VCC 1440 & 188.13912 & 15.41533 & Early & 8.83 & 16.0 & 0.99 & 0.72 & 4 & Y & 0.66 & C-M/L & -1.61 & 2.85 & 5.7 & 5 & 0.996 & L95 \\ 
VCC 1545 & 188.54808 & 12.04886 & Early & 8.85 & 16.83 & 1.01 & 1.12 & 3 & Y & 1.28 & C-M/L & -1.23 & 2.37 & 5.72 & 5 & 1.0 & L95 \\ 
VCC 1627 & 188.90521 & 12.38192 & Early & 8.8 & 15.63 & 1.12 & 1.14 & 3 & Y & 1.31 & C-M/L & -1.27 & 3.08 & 5.66 & 5 & 0.957 & L95 \\\relax 
[KK2000] 03 & 36.17792 & -73.51278 & Early & 8.16 & 2.0 & 0.0 & 0.93 & 1 & Y & 0.94 & C-M/L & -1.84 & 1.21 & 4.81 & 5 & 0.898 & H23 \\\relax 
[KK2000] 53 & 197.80917 & -38.90611 & Early & 6.85 & 2.92 & 0.0 & 1.01 & 1 & Y & 1.07 & C-M/L & -1.65 & 0.84 & 3.07 & 5 & 0.93 & H23 \\\relax 
[KK98] 096 & 162.61292 & 12.36083 & Early & 7.07 & 10.0 & 0.9 & 0.96 & 1 & Y & 0.98 & C-M/L & -2.31 & 1.1 & 3.36 & 5 & 0.939 & H23 \\ 
\enddata

\tablecomments{This table provides all relevant data for our sample galaxies. All galaxy masses (5), distances (6), and galaxy g-i colors (7) come from 50MGC \citep{Ohlson2023}, except for the masses of BTS 109, UGC 07242, [KK2000] 03, and [KK2000] 53. For these galaxies, we use masses from \citet{Hoyer2023a}, but found no alternative sources for galaxy-wide colors. Column (8) gives the nuclear colors within a r=$0\farcs5$ aperture, while (9) indicates the source of the nuclear color: 1=\citet{Hoyer2023a}, 2=\citet{Pechetti2020}, 3=aperture photometry, 4=SDSS psfMags. Column (10) indicates the if the galaxy is an NSC host. The i-band $M/L$'s are listed in column (11) with their source indicated in column (12): C-M/L=\citet{Taylor2011} color-$M/L$ relation, SED=$M/L$ derived from full spectrum SED modeling \citet{McDermid2015,Pechetti2017}. Columns (13) and (14) give the power-law fit parameters ($\gamma$=slope, $\rho_{5pc}$=3D stellar density at r=5pc) for the inner density profiles used to create the new scaling relations. Black hole masses used for the influence radius investigation are given in column (15) with the MBH mass source indicated in column (16): 1=\citet{van-den-Bosch2016}, 2=\citet{Reines2015}, 3=\citet{Greene2020}, 4=M$_{BH}$-$\sigma$ relation, 5=M$_{BH}$-M$_{gal}$ relation.\newtext{Column (17) gives the inner observed axial ratio used to define the spherically averaged density profile}. Lastly, column (18) indicates the parent sample for the surface brightness data.}

\end{deluxetable*}
\end{longrotatetable}

\end{document}